\documentclass[preprint,12pt,nofootinbib]{revtex4-1}

\usepackage{epsfig,amsmath,amsfonts,amssymb}
\usepackage{hyperref}

\begin{document}

\title{Boson dark matter halos with a dominant \\noncondensed component}

\author{Iskander G. Abdullin}
\email{videns42@gmail.com} \affiliation{Department of
	General Relativity and Gravitation, Institute of Physics, Kazan
	Federal University, Kremlevskaya str. 18, Kazan 420008, Russia}

\author{Vladimir A. Popov}
\email{vladipopov@mail.ru} \affiliation{Department of
	General Relativity and Gravitation, Institute of Physics, Kazan
	Federal University, Kremlevskaya str. 18, Kazan 420008, Russia}

\begin{abstract}
We consider galaxy halos formed by dark matter bosons with mass in the range of about a few tens or hundreds eV. A major part of the particles is in a noncondensed state and described under the Thomas--Fermi approach. Derived equations are solved numerically to find the halo density profile. 
The noncondensed state is supported in the entire halo except compact gravitationally bounded Bose--Einstein condensates. Although the size of these compact objects, also known as Bose stars, depends on interactions between the particles, its upper limit is only about 100 astronomical units. The Bose stars collect the condensed bosons providing a density cusp avoidance in the halo as well as a natural mechanism to prevent overproduction of small halos.
Clusters of the Bose stars can also contribute to the halo density profile.
The model is analyzed by confronting its predictions with observations of galaxy rotation curves. We employ 22~low surface brightness galaxies and obtain that the model is consistent with the observational data when the particle mass is in the range above about 50 eV and the best fit corresponds to the mass $m=86$~eV. This mass is appropriate for relic dark matter bosons, which decouple just after QCD phase transition.
\end{abstract}


\maketitle

\section{Introduction}

Observational data clearly indicate that about a quarter of the energy density in the Universe belongs to nonbaryonic dark matter (DM).
A number of theories provide us with various candidates for DM that are not included in the Standard Model of particle physics~\cite{Bertone2005Particle}.

Models, in which DM particles were nonrelativistic at decoupling from the thermal bath in the expanding Universe, are preferred from the cosmology standpoint. The particles of this kind form cold dark matter (CDM). Collisionless weakly interacting massive particles (WIMP) are the most favored candidates for CDM \cite{Mukhanov2005Physical}.
WIMPs naturally produce the modern residual DM density, and $N$-body simulations reproduce the main properties of large-scale structures \cite{Navarro1996structure}. Nevertheless, there are difficulties when applying to small scales.
A considerable challenge for this model is a central cusp in halo density profiles. The simulations show that close to the center of the halo the density behaves as $\rho\propto r^{-1}$ while observations indicate the more flattened profiles \cite{Bullock2017Small,Blok2010Core}. In addition, the predicted number of dwarf galaxies is much larger than it follows from observations \cite{Bullock2017Small,Strigari2007Redefining}.

Another type of CDM includes light (pseudo)scalar bosons forming a Bose--Einstein condensate (BEC) (see, e.g., \cite{Suarez2014review} and references therein). The condensed bosons have zero momentum and therefore contribute as a nonrelativistic component independently of their mass. An important advantage of BEC DM is that the central cusp in the halo density distribution is naturally eliminated because all the bosons are in the same quantum state and gravitational collapse is prevented by the Heisenberg uncertainty principle \cite{Hu2000Fuzzy}. An alternative way to solve the problems above within the BEC DM paradigm implies that repulsive self-interaction between the condensed particles balances gravity \cite{LeeBS,Boehmer2007Can}. This model establishes the maximum mass for particles capable to form DM halos as gigantic BEC. Due to astrophysical constraints on self-interaction the boson mass is less than about $m\sim 10^{-3}$~eV \cite{Slepian2012Ruling}.

The BEC DM approach covers particles with very different masses. The ultralight bosons have the mass $m\sim 10^{-22}$~eV or less, so that the de Broglie wavelength is comparable to the galactic size providing wave behavior on astrophysical scales. These particles have no need of self-interaction to create galaxy-sized BEC and therefore are governed by the Klein--Gordon--Einstein equations in the relativistic case~\cite{Seidel1990Dynamical} and the Schr\"odinger--Poisson equations for nonrelativistic behavior~\cite{Sin1994Late}. The scalar field without self-interaction with so tiny mass forms fuzzy CDM \cite{Schive2014Cosmic,Hu2000Fuzzy,Du2016Substructure}. Its feature is that the Heisenberg principle provides a minimum radius of gravitationally bound BEC, which can contribute to solving the missing satellites problem without disturbing the large-scale hierarchy \cite{Schive2014Cosmic,Press1990Single,Lee2010Minimum}.
However, as discussed in \cite{Safarzadeh2020Ultra}, the halo density profiles of the ultra-faint dwarf satellites appear to be incompatible with the ultralight DM.

Axions  and axion like particles (ALPs) are the most searched for CDM particles among the light bosons \cite{throuWall,ALPS,CAST,Caldwell2017Dielectric,ADMX}. The axion is a pseudoscalar boson  and a natural candidate for DM as a particle introduced in an extension of the Standard Model to solve the strong CP~problem through the Peccei--Quinn (PQ) mechanism \cite{PQ}. The PQ axion mass is defined by
\begin{equation}\label{eq:axion_mf}
m_\text{a}\sim 6\cdot \left( \frac{f_\text{a}}{10^{12}\ \text{GeV}}\right)^{-1} \mu\text{eV}\,,
\end{equation}
where $f_\text{a}$, an axion decay constant, determines the PQ symmetry-breaking scale. Experimental and astrophysical constraints exclude its values below $10^9$~GeV, so the axion mass is less than $10^{-2}$~eV \cite{Duffy2009Axions} whereas the cosmologically appropriate axion mass is of order $10^{-6}-10^{-5}$~eV \cite{Erken2012Axion}.

ALPs appear beyond the strong CP problem so that condition (\ref{eq:axion_mf}) is relaxed or evaded. There exist several mechanisms to decouple the pseudo-Goldstone boson mass from the decay constant. This idea has been embodied in compactification scenarios of the string theory \cite{Conlon2006The,Svrcek2006Axions,Arvanitaki2010String} or in the context of the relaxion mechanism \cite{Graham2015Cosmological,Espinosa2015Cosmological}. ALPs can have a wide range of implications for cosmology and astrophysics. Heavy ALPs are considered as mediators for interactions between the DM and Standard Model \cite{LAZARIDES2020135603} and also incorporated into SUSY DM models~\cite{Bae2015Mixed}. As regards sufficiently light ALPs, the pseudoscalars with mass below the MeV are very long-lived and may well constitute DM particles~\cite{Giannotti2011New,Arias2012WISPy}.

The particles with mass below $10^{-3}$~eV, including the PQ axions and ultralight bosons, are used to describe the DM halos as gigantic BECs \cite{Hu2000Fuzzy,LeeBS,Boehmer2007Can,Slepian2012Ruling,Seidel1990Dynamical,Sin1994Late,Schive2014Cosmic,Chavanis2021Jeans}. It is possible that sub-eV particles have to be excluded from this approach \cite{BEREZHIANI2021Core}. In this case the sub-eV particles can create only comparatively small BECs, otherwise recognized as Bose stars, and the halo is formed as a large field of BEC miniclusters \cite{Levkov2018Gravitational}. More massive (pseudo)scalars are inappropriate for the galaxy-sized BECs and therefore had previously escaped a high interest.  However, the particles with mass above eV are able to form simultaneously galaxy-sized objects when nondegenerate, and considerably smaller objects when degenerate. In this case the DM halo can involve two fractions, clustered Bose stars and noncondensed bosons.

We follow the semiclassical or Thomas--Fermi (TF) approach to examine the DM bosons within this configuration.  The TF approximation gives us a good way to take into account the quantum properties of DM particles inside the galaxy halos. In \cite{Destri2013Quantum,deVega2014Observational,deVega2017Equation} it was applied to fermionic warm DM (WDM) where quantum effects become apparent inside the halo core. The TF treatment is also valid for the noncondensed bosons near the critical point when a large fraction of particles occupy low quantum levels with small momenta. The TF approximation is also a part of the terminology for BEC~\cite{Griffin1996Conserving,Dalfovo1999Theory,pethick_smith_2008,Griffin2009Bose,Yukalov2011Basics}, including the BEC DM halos \cite{Boehmer2007Can,Chavanis2021Jeans,Chavanis2011Mass,Rindler-Daller2012Angular,Zhang2018Slowly,Guzman2014Rotation}, although it comes into use only because of resemblance in formulas~\cite{pethick_smith_2008}.

The paper is organized as follows. The basic properties of the BEC DM halos are reviewed in Section~\ref{sec:BEC}. In Section~\ref{sec:PCDM} we describe the DM halo consisting of noncondensed bosons.
A two-component model including both noncondensed bosons and condensed particles assembled into the Bose stars
is also considered.
Section~\ref{sec:RotationCurves} contains a comparison between model-predicted galactic rotation curves and observational data from low surface brightness (LSB) galaxies. Finally, we discuss and conclude our results.

\section{BEC DM halos}
\label{sec:BEC}

\subsection{Pure condensed DM}

The nonrelativistic BEC DM is generally represented as a complex scalar field $\psi(\mathbf r)$ with repulsive self-interaction. In virialized halos its distribution is described by the time-independent Gross-Pitaevskii equation \cite{Boehmer2007Can,Chavanis2011Mass,Rindler-Daller2012Angular}
\begin{equation}\label{eq:GPeq}
\left(-\frac{\hbar^2}{2m}\nabla^2 + m \Phi(\mathbf r) + gn(\mathbf r) - \mu \right) \psi(\mathbf r) =0\,,
\end{equation}
where $n(\mathbf r)=|\psi(\mathbf r)|^2$ is identified as the number density of particles with mass $m$, $\mu$ is a chemical potential, $g$ is a coupling constant governing the interaction between the particles, and the  gravitational potential $\Phi(\mathbf r)$ obeys the Poisson equation
\begin{equation}\label{eq:Poisson}
\nabla^2 \Phi(\mathbf r)=4\pi G m n(\mathbf r)\,.
\end{equation}
The coupling constant can be also expressed in terms of the $s$-wave scattering length $a$
\begin{equation}\label{eq:g}
g=\frac{4\pi \hbar^2 a}{m}\,,
\end{equation}
so that the scattering cross-section is quadratic in $g$
\begin{equation}\label{eq:crosssec_via_g}
\sigma=8\pi a^2 =\frac{m^2g^2}{2\pi\hbar^4}\,.
\end{equation}

Eq.~(\ref{eq:GPeq}) implies balance between gravity and effective pressure involving two terms \cite{Chavanis2011Mass}. The first term, so-called quantum pressure, corresponding to the kinetic term in Eq.~(\ref{eq:GPeq}) has its anisotropic components
\begin{equation}\displaystyle
(P_\text{q})_{ij} = -\frac{\hbar^2}{4m^2}\left(\nabla^2\psi(\mathbf r)\delta_{ij} - 4\partial_i\psi(\mathbf r)\partial_j \psi(\mathbf r)\right)\,,
\end{equation}
while the second one is provided by the repulsive interaction between the particles
\begin{equation}
\displaystyle P_\text{int} = \frac12 g n^2(\mathbf r)\,.
\end{equation}

The quantum pressure is taken into account when the de Broglie wavelength is comparable to a BEC extent.
For galaxy-sized BEC the appropriate de Broglie wavelength is found as
\begin{equation}
\lambda_\text{dB}\sim\frac{\hbar}{m v}\,,
\end{equation}
where $v^2$ being a velocity dispersion of the DM bosons. The corresponding boson mass is estimated according to
\begin{equation}\label{eq:massULB}
m\lesssim\frac{\hbar}{\lambda_\text{dB}\, v}\sim 1.9\cdot 10^{-23} \,\left(\frac{\lambda_\text{dB}}{1 \text{\ kpc}}\right)^{-1}  \left(\frac{v}{100 \text{\ km/s}}\right)^{-1} \ \text{eV}\,.
\end{equation}

For the scalar particles falling under condition (\ref{eq:massULB}) the quantum pressure is always taken into consideration while the self-interaction can be ignored. This approximation is recognized as a \emph{kinetic regime}. The ultralight bosons without self-interaction form the fuzzy DM, which demonstrates wave nature on the galaxy scales.  Its wave properties suppress the central cusp in the DM halos and reduce the abundance of low mass halos \cite{Hu2000Fuzzy,Lee2010Minimum}  whereas the large scale structure is similar to CDM although galaxy formation can be delayed relative to CDM \cite{Schive2014Cosmic}.

The de Broglie wavelength for nonrelativistic axions can be estimated as
\begin{equation}\label{eq:dBaxions}
\lambda_\text{dB}\sim 6\cdot 10^{4} \,\left(\frac{m}{10^{-6} \text{\ eV}}\right)^{-1}  \left(\frac{v}{100 \text{\ km/s}}\right)^{-1} \ \text{cm}\,.
\end{equation}
This scale implies that DM consisting of non-interacting axions can break down into relatively small condensates identified as axion stars, which form a cluster structure of the halo \cite{Hogan1988Axion,Levkov2018Gravitational,Eggemeier2019Formation}. The size of the axion stars can be much greater than estimation (\ref{eq:dBaxions}) if the particle velocity turns out to be very small \cite{Sikivie2009Bosea,Mielczarek2010Vortex}. The upper limit can be obtained immediately from Eq.~(\ref{eq:GPeq}) when the first two terms are of the same order,
\begin{equation}\label{eq:AxStarR}
R_c\lesssim\left(\frac{3\hbar^2}{8\pi G m^2\rho}\right)^{1/4}\sim 1.9\,\left(\frac{m}{10^{-6} \text{\ eV}}\right)^{-1/2}  \left(\frac{\rho}{10^{-24} \text{\ g/cm}^3}\right)^{-1/4} \ \text{\ a.\,u.}\,,
\end{equation}
where $\rho\sim 10^{-24}\text{\ g/cm}^3$ corresponds to  the halo density although the axion stars are certainly denser.

The opposite approach is used for the scalar particles with masses being out of inequality (\ref{eq:massULB}). On the galactic scales the kinetic term is negligible and the equilibrium distribution is supported by the repulsive self-interaction. This approach is known as the \emph{TF approximation}\footnote{Eq.~(\ref{eq:GPeq}) without the kinetic term is reminiscent of relation for the Fermi energy in the TF approximation in the theory of atoms, so the approximation for BEC is generally referred to by the same name.}.\label{page:footnote1} It is well suited for both spatially homogeneous BECs and inhomogeneous systems in trapping potentials  \cite{Griffin1996Conserving,Dalfovo1999Theory,pethick_smith_2008,Griffin2009Bose,Yukalov2011Basics}.

The self-interacting DM naturally solves the cuspy halo problem \cite{Tulin2018Dark}, can explain the Tully-Fisher relation \cite{Mo2000Tully}, and is also compatible with DAMA experiment \cite{Mitra2005Has}. Nevertheless, the interaction strength is constrained by the cosmological grounds \cite{bernal2019phenomenology,Robertson2019mnras} and the observations of colliding galaxies. According to the measurements for the Bullet Cluster \cite{Randall2007Constraints} and the cluster MACS J0025.4-1222 \cite{Bradac2008Revealing} the cross-section to mass ratio is estimated as
\begin{equation}\label{eq:crosssec_to_mass}
\begin{array}{l}
\displaystyle \frac{\sigma}{m}<1.25\ \text{cm}^2\cdot\text{g}^{-1} \quad \text{(Bullet Cluster)}\,, \\[12pt]
\displaystyle\frac{\sigma}{m}<4\ \text{cm}^2\cdot\text{g}^{-1} \quad \text{(MACS J0025.4-1222)}\,.
\end{array}
\end{equation}
The corresponding constraints for interaction follow from relation (\ref{eq:crosssec_via_g}).

A spherically symmetric solution to Eqs. (\ref{eq:GPeq}) and (\ref{eq:Poisson}) in the TF approximation read as
\begin{equation}\label{eq:BECdens}
n(r)=n_0\,\frac{\sin(\pi r/R_c)}{\pi r/R_c}\,,
\end{equation}
where $n_0=n(0)$ is a central particle density, and
\begin{equation}\label{eq:Rc}
R_c=\sqrt{\frac{\pi^2 g}{4\pi G m^2}} =\sqrt{\frac{\pi^2\hbar^2 a}{G m^3}}
\end{equation}
is the BEC radius, which is assumed to be comparable with the halo size \cite{Boehmer2007Can}.

Combining (\ref{eq:g}), (\ref{eq:crosssec_to_mass}) and (\ref{eq:Rc}) entails the constraint for the mass of the self-interacting condensed particles \cite{Slepian2012Ruling}
\begin{equation}\label{eq:mass_estim_1}
m\lesssim\left(\frac{(\sigma/m)\pi^3\hbar^4}{8 G^2 R_c^4}\right)^{1/5} \sim 9.6\cdot 10^{-4} \left(\frac{R_c}{1 \text{\ kpc}}\right)^{-4/5} \left(\frac{\sigma/ m}{1.25\ \text{cm}^2/\text{g}}\right)^{1/5} \ \text{eV}\,.
\end{equation}

It is clear that the DM axions satisfy this condition and can be considered as a self-interacting DM candidate. The axion scattering length in this case is restricted by
\begin{equation}\label{eq:axion_scat_length}
a\lesssim 9.4\cdot 10^{-21}\left(\frac{\sigma/ m}{1.25\ \text{cm}^2/\text{g}}\right)^{1/2}\left(\frac{m}{10^{-6} \text{\ eV}}\right)^{1/2}  \ \text{cm}\,.
\end{equation}

Constraint (\ref{eq:crosssec_to_mass}) can be stronger if there is the Bose enhancement of the collision process. In this case the cross-section to mass ratio in (\ref{eq:crosssec_to_mass}) is multiplied by the large occupation number of the degenerate particles, and (\ref{eq:mass_estim_1}) reduces to $m\lesssim 10^{-15}$~eV \cite{BEREZHIANI2021Core}.

Solution (\ref{eq:BECdens}) implies that gravitationally bound self-interacting BECs have the same radius independently of their mass. Clearly, it is unacceptable for realistic DM halos. The extended BEC DM models take into account available baryonic matter \cite{Dwornik2017Bose}, rotation \cite{Zhang2018Slowly,Rindler-Daller2012Angular,Guzman2014Rotation} and  temperature \cite{Harko2012Finite,Slepian2012Ruling,Abdullin2019Bose} effects. In these approaches $R_c$ is identified as a minimum radius corresponding to an inner halo core while the halo can be well larger.

The models of BEC DM halos implies a high macroscopic occupancy of the many-body ground state, which takes place when thermalized bosons are very dense, i.\,e. $n\lambda_\text{dB}^3\gg 1$. This condition is equivalent to there being a critical temperature, below which BEC can be formed. In the self-interacting BEC DM models the interaction can be responsible for relaxation towards a statistical equilibrium. In this case the interaction is expected to be sufficiently strong; otherwise, the corresponding relaxation timescale at galactic densities is over the age of the Universe \cite{Slepian2012Ruling}. In order for the bosons to thermalize and form BEC via the self-interaction, the relaxation time, $t_\text{rel}^{-1}\sim n\sigma\bar v$ (here $\bar v $ is the mean particle velocity), is to be less than the Hubble time $t_\text{H}\sim H^{-1} \approx 4.6\cdot 10^{17}$~sec. Besides, eliminating the cross section in favor of the BEC radius by (\ref{eq:crosssec_via_g}) and (\ref{eq:Rc}) this requirement implies the particle mass 
\begin{equation}
m \gtrsim \left(\frac{\pi^3\hbar^4 H}{8G^2 R_c^4\rho\bar v}\right)^{1/5} \sim 6.8\cdot 10^{-4}  
\left(\frac{R_c}{1 \text{\ kpc}}\right)^{-4/5} 
\left(\frac{\rho}{10^{-24} \text{\ g/cm}^3}\right)^{-1/5}
\left(\frac{\bar v}{100 \text{\ km/s}}\right)^{-1/5}
\ \text{eV}\,.
\end{equation}
Thus, the self-interacting DM bosons creating galaxy-sized BECs ought to possess mass very close to 10$^{-3}$~eV in this case.

On the other hand, this limitation can be obviated by the concept of gravitational cooling, first introduced in \cite{Seidel1994Formation} for the problem of scalar field perturbations collapsing into Bose stars. This process is similar to violent relaxation \cite{Lynden-Bell1967Statistical} developed to explain the rapid relaxation of collisionless particles. The cooling mechanism also provides the relaxation time much shorter than $t_\text{rel}$ and can be applied to both the interacting and non-interacting DM bosons \cite{Guzman2006Gravitational,Chavanis2019Predictive}. Thus, even though the self-interaction is still needed, say, to stabilize the BEC core, the likely scenario for halo evolution would comprise gravitational cooling and violent relaxation, which allow the boson DM halos to reach the equilibrium configuration avoiding the ``collisional'' restrictions \cite{Chavanis2019Predictive}.

\subsection{BEC DM with a nondegenerate component}
\label{sec:BEC-nc}

It seems proper to take always into account the nondegenerate component  when describing BEC at nonzero temperature. In this case the halo in thermal equilibrium involves a core, in which some bosons form BEC, while the others remain in the noncondensed state. Outside the core the bosons are in short supply to be condensed so the core is surrounded by a cloud of the noncondensed bosons. The nondegenerate component is incorporated into consideration by decomposition of the scalar field into the condensed and noncondensed parts, $\psi=\psi_\text{c}+\psi_\text{n}$. Accordingly, the particle density in Eq.~(\ref{eq:Poisson}) is represented as $n=n_\text{c}+n_\text{n}$ while Eq.~(\ref{eq:GPeq}) in the TF approximation read as
\begin{equation}
m\Phi + gn_\text{c} + 2gn_\text{n} + gs_\text{n} = \mu\,,
\end{equation}
where $s_\text{n}$ is an off-diagonal (anomalous) density resulting from the product $\psi_\text{n}\psi_\text{n}$ instead of $\psi_\text{n}^\dag \psi_\text{n}$ as for the normal density. The noncondensed and anomalous densities are self-consistently expressed via $n_\text{c}$ and $\Phi$ using the Bogoliubov transformations \cite{Yukalov2011Basics,Abdullin2019Bose}.

The BEC DM with the nondegenerate component was studied within the semi-classical \cite{Harko2012Finite,Slepian2012Ruling} and Hartree--Fock--Bogoliubov \cite{Abdullin2019Bose} approximations. It was shown numerically that in the galaxy-sized BEC core the noncondensed component gives only an insignificant contribution to the DM density profile and rotation curves when confronted with the pure BEC DM model.

At very low temperatures  the nondegenerate fraction into the DM density is of order of the condensate depletion parameter $(a^3n_0)^{1/2}$. It is assumed to be small for a dilute gas, and it is very negligible under condition (\ref{eq:axion_scat_length}) for all kinds of particles.

For higher temperatures there are two dimensionless parameters, which are responsible for the contribution of the nondegenerate component. The first one,
\begin{equation}\label{eq:nu}
\nu = \lambda_\text{dB}^3 \,n_0\,,
\end{equation}
is the particle number in the cubical volume of the de Broglie wavelength. In the condensed state the value of $\nu>\zeta(3/2)\approx 2.61$, where $\zeta$ is the Riemann zeta function. Its inverse value, $\zeta(3/2)/\nu$, measures the noncondensed fraction in the total particle density.
The parameter $\nu$ is independent of the interaction, and therefore can be employed to both the self-interacting and non-interacting particles.
The second parameter, $a/(\lambda_\text{dB}\nu)$,  is applicable only to the interacting bosons and affects the slope and shape of the DM density profile.

Keeping in mind estimation (\ref{eq:mass_estim_1}) one can easy obtain for the interacting particles
\begin{equation}\label{eq:inverse_nu}
\nu^{-1}\lesssim\frac{m^4 v^3}{\hbar^3\rho} \sim 7.3\cdot 10^{-18} \left(\frac{m}{9.6\cdot 10^{-4} \text{\ eV}}\right)^{4} \left(\frac{v}{100 \text{\ km/s}}\right)^{3} \left(\frac{\rho}{10^{-24} \text{\ g/cm}^3}\right)^{-1}\,,
\end{equation}
while the second parameter is even less since $a\ll\lambda_\text{dB}$.
Thus the nondegenerate fraction is unable to contribute noticeably to the total galaxy mass and has to be ignored in practical calculations.
It is all the more true for the ultralight non-interacting particles due to the fourth power of $m$ in estimation (\ref{eq:inverse_nu}).

A significant contribution from the nondegenerate component might be expected if $\nu$ is at least in the range near 10 (and it increases when $\nu$ come close to the value $\zeta(3/2)$ at the condensation point). In this case the bosons had the mass
\begin{equation}\label{eq:mass_estim_2}
m \gtrsim 10.4 \left(\frac{v}{100 \text{\ km/s}}\right)^{-3/4} \left(\frac{\rho}{10^{-24} \text{\ g/cm}^3}\right)^{1/4}\ \text{eV}\,.
\end{equation}
Clearly, it is comes into conflict with estimation (\ref{eq:mass_estim_1}), which rules out the nondegenerate component for  the gigantic BEC DM halos \cite{Slepian2012Ruling,Abdullin2019Bose}.

\section{Noncondensed DM halos}
\label{sec:PCDM}

\subsection{Pure noncondensed DM}

The contradiction between mass estimations (\ref{eq:mass_estim_1}) and (\ref{eq:mass_estim_2}) can be resolved when one considers BECs smaller in size than the galaxy extent. Even small-sized ``drops'' of the condensed particles, that is to say, the Bose stars, can be abundant in the halo and contribute significantly to the total halo mass. This regime will be considered below. Here we are concerned with a completely noncondensed DM halo. 
This regime is valid when BECs are few and far between in the halo, so the contribution of BECs is neglected, and $n(\textbf{r})$ in Eq.~(\ref{eq:Poisson}) now implies the noncondensed particle density $n_\text{n}(\textbf{r})$. In this case the noncondensed bosons can be described in the TF (or semiclassical) approximation\footnote{We remind, that the terminology  \emph{TF approximation} is also used in a somewhat different context for BEC (see remark on page~\pageref{page:footnote1}).}. It means that ensemble-averaged quantities are found as integrals over the momentum space with the Bose-Einstein distribution function
\begin{equation}\label{eq:distribBE}
f(\textbf{p},\textbf{r}) = \frac{1}{(2 \pi \hbar)^3}\,\frac{1}{e^{(\epsilon(\textbf{p},\textbf{r})-\mu)/kT}-1} \, ,
\end{equation}
where $k$ is the Boltzmann constant and $T$ is effective temperature, so that
\begin{equation}\label{eq:dispV0}
V_0^2=\frac{kT}{m}
\end{equation}
is the one-dimensional velocity dispersion. The particle energy in the TF approximation read as
\begin{equation}
\epsilon(\textbf{p},\textbf{r})=\frac{\textbf{p}^2}{2m}+m \Phi(\textbf{r}) + 2 g  n_\text{n}(\textbf{r})\,.
\end{equation}
It includes the mean-field potential providing by the self-interaction in addition to the first two conventional terms corresponding to the kinetic and potential energy \cite{Griffin2009Bose}.

The particle density is given by
\begin{equation}\label{eq:nonconddens}
n_\text{n}(\textbf{r}) = \int f(\textbf{p},(\textbf{r})) d^3 \textbf p = \lambda_\text{dB}^{-3}\, g_{3/2}\left( z(\textbf{r})\right)\,,
\end{equation}
where the de Broglie wavelength is determined as
\begin{equation}\label{eq:deBroglie}
\lambda_\text{dB} = \sqrt{\frac{2\pi\hbar^2}{mkT}} =  \frac{\sqrt{2\pi}\,\hbar}{m V_0}\,,
\end{equation}
and $g_{3/2}(z)$ is a special case of the polylogarithm functions $\displaystyle g_s(z)=\sum\limits_{i=1}^\infty \frac{z^i}{i^s}$ (sometimes $g_{3/2}(z)$ is also recognized as the Bose--Einstein function), and its argument in (\ref{eq:nonconddens}) is
\begin{equation}\label{eq:z}
z(\textbf{r}) = e^{(\mu - m \Phi(\textbf{r}) - 2 g  n_\text{n}(\textbf{r}))/kT}.
\end{equation}

Applying the nabla operator to (\ref{eq:nonconddens}) and using Eq.~(\ref{eq:Poisson}) one obtains
\begin{equation}\label{eq:ncEq}
\left(1 + \frac{2g }{\lambda_\text{dB}^{3}kT}\, g_{1/2}(z)\right)\nabla^2 n_\text{n}(\textbf{r}) - \lambda_\text{dB}^3\, \frac{g_{-1/2}(z)}{g_{1/2}(z)^2}\Bigl(\nabla n_\text{n}(\textbf{r})\Bigr)^2 + \frac{4 \pi G m^2}{ \lambda_\text{dB}^3kT}\, g_{1/2}(z)  n_\text{n}(\textbf{r}) = 0\,.
\end{equation}
Eqs.~(\ref{eq:ncEq}) and (\ref{eq:z}) describes the TF structure for the nondegenerate bosonic DM halos.

As it is expected for the nondegenerate bosons, self-interaction has not a key role as it takes place in BEC. Its contribution in the first term of (\ref{eq:ncEq}) is of order of $a/(\lambda_\text{dB}\nu) \ll 1$ and can be neglected without damage to further calculations.

Eq.~(\ref{eq:ncEq}) can be transformed into the dimensionless equation
\begin{equation}\label{eq:Bose-gas}
\nabla_\xi^2 x  - \nu\,\frac{g_{-1/2}(z)}{g_{1/2}(z)^2}(\nabla_\xi x )^2 + \frac{1}{\nu}\,g_{1/2}(z)x  = 0\,,
\end{equation}
where $x=n_\text{n}/n_0$ is a density normalized to its central value $n_0=n_\text{n}(0)$, $\xi= r/R$ is the dimensionless radial coordinate, and the scale radius
\begin{equation}\label{eq:inhR}
R^2 = \frac{kT}{4 \pi G m^2 n_0} = \frac{V_0^2}{4 \pi G m n_0}
\end{equation}
is the same as in the isothermal sphere (IS) model. Note, that (\ref{eq:inhR}) can be represented as $R^2=R_0^2/\nu$, where
\begin{equation}\label{eq:ncR}
R_0^2 = \frac{\lambda_\text{dB}^3kT}{4 \pi G m^2 } 
\end{equation}
is governed only by the effective temperature and independent on the central density. The parameter $\nu$ is defined as in (\ref{eq:nu}) and the quantity $z$ in Eq.~(\ref{eq:Bose-gas}) is found as a solution to the equation
\begin{equation}
\nu x = g_{3/2}(z)\,.
\end{equation}

\begin{figure}
	\centering
	\includegraphics[width=0.7\linewidth]{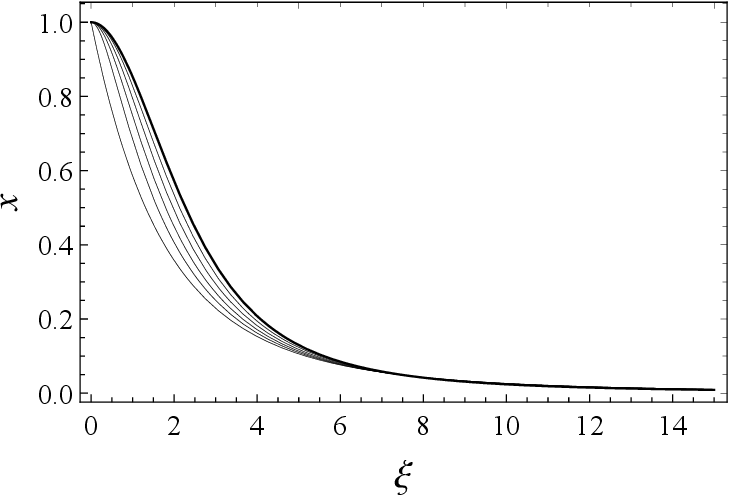}
	\caption{The density profiles for the completely noncondensed DM halo for $\nu = $ 0.5, 1, 1.5, 2, 2.61 (the upper to lower thin lines). The thick line corresponds to the IS model. }
	\label{fig:densProfile}
\end{figure}

Eq.~(\ref{eq:Bose-gas}) contains the only parameter $\nu$. Since the equation describes the nondegenerate bosons, the parameter $\nu$ lies in the range between zero and $\zeta(3/2)$. Values above this range correspond to the condensation region where Eq.~(\ref{eq:Bose-gas}) is not valid. The solutions to Eq.~(\ref{eq:Bose-gas}) are shown in Fig.~\ref{fig:densProfile}.

If the parameter $ \nu \ll 1$ the bosons are far from the condensation point and can be considered as a Maxwell-Boltzmann gas. In this case Eq.~(\ref{eq:Bose-gas}) is reduced to
\begin{equation}\label{eq:ISM}
\nabla_\xi^2 x  - \frac{1}{x}(\nabla_\xi x )^2 + x^2  = 0\,,
\end{equation}
corresponding to the IS model.

The IS profile is also reproduced by Eq.~(\ref{eq:Bose-gas}) at large $\xi$ when $x\ll 1$.
The same holds true for fermionic TF halos  \cite{deVega2014Observational}. This similarity in behavior shows that the particle statistics is crucial inside the compact galaxy core while the DM comes to a diluted regime corresponding to a classical gas outside the core.
There is no general agreement  what is to be considered as the halo core. Usually it is thought as a region, inside which the density decreases fixed number of times. In \cite{deVega2014Observational} this ratio is taken to be four in accordance with the empirical Burkert profile. Following the same choice one finds from  Eq.~(\ref{eq:Bose-gas}) the dimensionless core radius $\xi_\text{core}$ depending on $\nu$.  The corresponding dimensional radius is represented as
\begin{equation}
R_\text{core} = \frac{\xi_\text{core}}{\sqrt\nu}R_0\,.
\end{equation}

\begin{figure}
\centering
\includegraphics[width=0.7\linewidth]{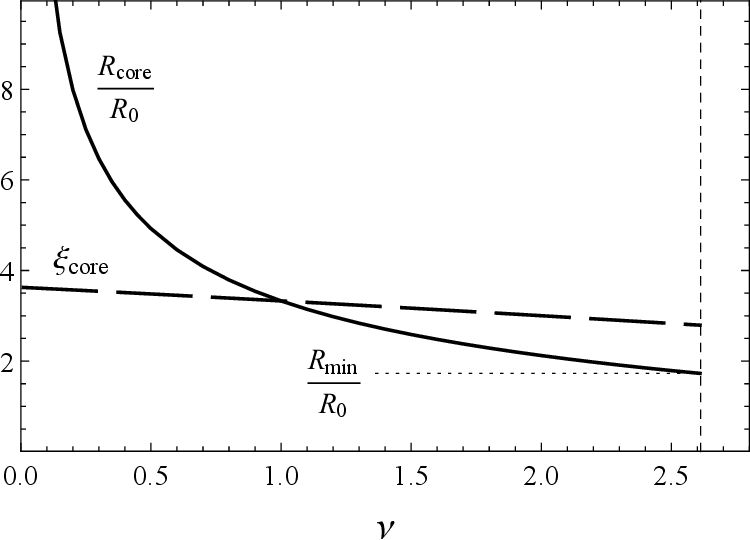}
\caption{The dimensionless core radius $\xi_\text{core}$ (dashed line) and the corresponding dimensional radius in units of $R_0$ (solid line) depending on the parameter $\nu$. The thin vertical dashed line marks the critical value $\nu=\zeta(3/2)$. }
\label{fig:cutR}
\end{figure}

It is seen in Fig.~\ref{fig:cutR} that the core radius $\xi_\text{core}$ is nearly unvaried so that at fixed temperature $R_\text{core}$ decreases as the parameter $\nu$ increases, and reaches its minimum $R_\text{min}$ at the condensation point when $\nu=\zeta(3/2)$. 
Below this threshold, the central particle density is large enough for the condensed bosons to be available inside the halo.
The condensed bosons are assembled into compact BECs while the surrounding particles remain noncondensed and provide the core radius $R_\text{core}\geqslant R_\text{min}$. The BEC radius $R_c \ll R_\text{min}$, and the condensed particles can build up galaxy-sized objects only through clustering of isolated BECs. This mechanism favors halo formation above the threshold radius and can be considered as an appropriate explanation why the smaller-sized galaxies are in a deficiency while a sufficient number of the larger-sized ones is observed.

\subsection{Bose stars in noncondensed DM halos}
\label{sec:BS}

Taking into account restrictions (\ref{eq:crosssec_to_mass}) and (\ref{eq:mass_estim_2}) one obtains the BEC radius for the self-interacting particles
\begin{equation}\label{eq:RbsSIDM}
R_c=\left( \frac{\left(\sigma/ m\right) \hbar^4}{8\pi G^2 m^5} \right)^{1/4} \lesssim
83.7\, \left(\frac{\sigma/ m}{1.25\ \text{cm}^2/\text{g}}\right)^{1/4} \left(\frac{m}{50\ \text{eV}}\right)^{-5/4} \text{\ a.\,u.}
\end{equation}

As it is shown in \cite{Abdullin2019Bose} the density of the nondegenerate component in self-gravitating BEC is practically invariable so that the BEC density distribution is described by Eq.~(\ref{eq:BECdens}) with an additional constant term corresponding to the noncondensed background. According to (\ref{eq:RbsSIDM}) the relatively small compact objects, the Bose stars, can be found within a halo consisting of the noncondensed particles. However, details of the BEC density profiles are unessential for the halo structure and are beyond the scope of our consideration.

The Bose stars can arise from overdensities in the inner regions during DM halos virialization. Another way implies the production of small-scale DM clumps from large amplitude perturbations on corresponding scales in the early Universe \cite{Kolb1994Large,Berezinsky2003Small}. The zero momentum condensed particles are collected in gravitational wells and form compact BECs even at the background DM density \cite{Breit1984Cold,Seidel1994Formation} while the clumps can be  considerably denser \cite{Kolb1994Large} making possible clusters of Bose stars. The relatively small clusters, or BEC miniclusters for short, are expected to be arranged in the halo accumulating mostly in the core. Large groups of the Bose stars are capable to form quite massive gravitationally bound objects comparable with
compact dwarf galaxies.

The BEC miniclusters can play a twofold role in the halo. On the one hand, the Bose stars provide an appropriate way to create a barrier for cuspy density profiles.  During the structure formation, the density fluctuations on small scales make favorable conditions for Bose stars production. Gravitationally bounded compact BECs are formed around density peaks while surrounding regions filled with noncondensed particles remain underdense, preventing the steep density growth on larger scales.

On the other hand, the aggregate mass of the BEC miniclusters can be enough to be appreciable in the total halo mass and thereby contribute to the halo density profile.
An inhomogeneous structure because of the BEC miniclusters can be quite important for methods of DM detection. The effects of clumpiness on signals from DM of various nature were studied in \cite{Kolb1994Large,Bergstrom1999Clumpy,Berezinsky2003Small}. However, there is no accurate knowledge of the influence of DM clumps inside the halo on observed kinematic effects. Allowing a possibility for the certain DM fraction to be condensed with high local density, we consider its spatial distribution to be smooth on the galaxy scales. The corresponding density profile should be regarded as a function, which describes an average distribution of the BEC miniclusters in the galactic halo, the standard approach to describe a large number of stars in a galaxy.

A fundamental description of this system is specified by a distribution function, which obeys the collisionless Boltzmann equation. To single out the spatial properties of the system one can use the reduced representation by taking the moments of the Boltzmann equation.

The first moment is recognized as the Jeans equation.
For an equilibrium spherical system, assuming the velocity anisotropy to be vanish, it reads
\begin{equation}\label{eq:Jeans}
\nabla(v^2\rho_\text{bs}) + \rho_\text{bs}\nabla \Phi = 0\,,
\end{equation}
where $\rho_\text{bs}$ is a mass density, and $v^2$ is a one-dimensional velocity dispersion for the Bose stars.
In form (\ref{eq:Jeans}) the Jeans equation reproduces the hydrostatic equilibrium condition for an ideal gas with the pressure $p_\text{bs}=v^2\rho_\text{bs}$ and provides the IS density profile for the BEC fraction of DM.

The right-hand side of Eq.~(\ref{eq:Poisson}) for the gravitational potential $\Phi$ now incorporates two parts corresponding the noncondensed particles and the Bose stars
\begin{equation}
\nabla^2 \Phi(\mathbf r)=4\pi G(mn_\text{n}(\mathbf r)+\rho_\text{bs}(\mathbf r))\,,
\end{equation}
where the noncondensed particle density $n_\text{n}$ is subject to Eqs.~(\ref{eq:nonconddens}) and (\ref{eq:z}) as before.

Using the same normalized quantities as in Eq.~(\ref{eq:Bose-gas}) we obtain the dimensionless equations for two-component DM consisting of the Bose stars and the noncondensed particles,
\begin{equation}\label{eq:2components}
\begin{array}{l}
\displaystyle
\nabla_\xi^2 x  - \nu\,\frac{g_{-1/2}(z)}{g_{1/2}(z)^2}(\nabla_\xi x )^2 + \frac{1}{\nu}\,g_{1/2}(z)(x+x_\text{bs})  = 0\,,
\\ \displaystyle
\nabla_\xi x_\text{bs} = \gamma \nu\frac{ x_\text{bs} }{g_{1/2}(z)} \nabla_\xi x\,.
\end{array}
\end{equation}
For the sake of simplicity, we take the velocity dispersion of the Bose stars to be constant so that it is involved in the parameter $\gamma=V_0^2/v^2$, which also specifies the scale length of the Bose stars distribution. This scale, on the one hand, can be defined as usual \cite{binney2011galactic}
\begin{equation}
R_\text{bs}^2 = \frac{v^2}{4\pi G\rho_\text{bs}(0)}\,.
\end{equation}
On the other hand, according to the second equation in (\ref{eq:2components}),
\begin{equation}
R_\text{bs}^2 \sim \frac{R^2/\nu}{\gamma^2} = \frac{1}{\gamma^2}\,\frac{V_0^2}{4\pi G m n_\text{n}(0)}\,,
\end{equation}
and hence
\begin{equation}\label{eq:RatioCentrDens}
\alpha=\frac{\rho_\text{bs}(0)}{m n_\text{n}(0)}\sim \gamma\,.
\end{equation}
This relation implies a steeper density profile for the DM component in Eqs.~(\ref{eq:2components}), which proves to be dominant in the center of the halo.
Thus, the BEC miniclusters accumulate in the central core when in abundance, and are almost uniform in the halo when in a small quantity. In the next section Eqs.~(\ref{eq:2components}) are used  to reproduce observed rotation curves when the pure noncondensed model alone is unsuitable for a good fit.

The considered approach can be also applied to the non-interacting bosons. In this case the Bose stars are formed similar to the axion stars in the kinetic regime \cite{Levkov2018Gravitational}. However, the Bose stars are much smaller than estimation (\ref{eq:AxStarR}) owing to the larger particle mass,
\begin{equation}
R_c\lesssim 4\cdot 10^{9}\,\left(\frac{m}{50 \text{\ eV}}\right)^{-1/2}  \left(\frac{\rho}{10^{-24} \text{\ g/cm}^3}\right)^{-1/4} \ \text{cm}\,.
\end{equation}

There is a more fundamental difference in the halo structure between these kinds of particles. The axion miniclusters are the sole DM component in the halos because the nondegenerate part is ignorable while in our model the BEC miniclusters are surrounded by the noncondensed bosons, so both fractions can contribute to the total halo mass.

\subsection{Constraints on the boson mass}

Here we discuss an appropriate particle mass range, which is able to provide simultaneous contributions of both noncondensed and condensed bosons to the halo density profile. As it was reviewed in Section~\ref{sec:BEC-nc}, the noncondensed DM halo is provided by bosons with mass over 10~eV  while well-motivated nonthermal mechanisms for axions and ALPs imply quite smaller \cite{throuWall,ALPS,CAST,Caldwell2017Dielectric,ADMX,Duffy2009Axions,Erken2012Axion} or larger \cite{LAZARIDES2020135603,Bae2015Mixed,Giannotti2011New,Arias2012WISPy} particle mass. Keeping open a theoretical possibility of nonthermal production, we now address to thermal relic DM bosons.

An upper bound on the particle mass follows from the estimation of the present abundance of the DM bosons including BEC \cite{Boyanovsky2008Constraints}
\begin{equation}\label{eq:DMabundance}
\rho_\text{DM} = \rho_\text{n} \left(\frac{T_\text{c}}{T_\text{d}}\right)^3 =\frac{2\zeta(3)m}{\pi^2g^*_\text{d}}\left(\frac{kT_0}{\hbar c}\right)^3\left(\frac{T_\text{c}}{T_\text{d}}\right)^3  \,,
\end{equation}
where $\rho_\text{n}$ is the energy density of the nondegenerate bosons, $T_0$ is the cosmic microwave background temperature today,  $T_\text{c}$ and $T_\text{d}$ are the condensation and decoupling  temperatures respectively ($T_\text{c}\geqslant T_\text{d}$), $c$ is the light velocity, and $g^*_\text{d}$ is the number of relativistic degrees of freedom at decoupling.

Expression (\ref{eq:DMabundance}) is obtained for the light bosons that decoupled in thermal equilibrium while the nondegenerate component was relativistic at the temperature $kT_\text{d}\gg mc^2$, and the particle number density $n_\text{n}\propto T^3$. After recombination, when the bosons become nonrelativistic, this dependence is also kept since the distribution is frozen for the particles decoupled from the thermal bath. The corresponding present energy density is represented as usual
\begin{equation}\label{eq:ncEnergy}
\rho_\text{n}(t) = m c^2 n_\text{n}(t)\,,
\end{equation}
providing the third power of temperature in (\ref{eq:DMabundance}).

The condensed fraction is always nonrelativistic and evolves as CDM immediately after decoupling producing the increase factor $\left(T_\text{c}/T_\text{d}\right)^3$ with respect to the density of the completely noncondensed DM.

Keeping in mind that the light bosons can be only a part of DM, one gets the constraint
\begin{equation}\label{eq:mUpper}
m\lesssim 0.2 \left(\frac{\Omega_\text{DM}}{0.25}\right) \left(\frac{g^*_\text{d}}{61.75}\right) \left(\frac{T_\text{d}}{T_\text{c}}\right)^3~\text{keV}\,,
\end{equation}
where $\Omega_\text{DM}$ is the share of DM in the present Universe energy density.

The lower mass bound in our model emerges naturally from the condition $\nu\leqslant\zeta(3/2)$. It determines the threshold core radius, below which bosons cannot remain noncondensed at the center of the halo and BEC is formed. Substituting (\ref{eq:deBroglie}) and (\ref{eq:ncEnergy}) in (\ref{eq:nu}) one finds that
\begin{equation}\label{eq:mLowerBEC}
m\gtrsim  \left( \frac{(2\pi)^{3/2}}{\zeta(3/2)} \frac{\hbar^3\rho(0)}{V_0^3} \right)^{1/4} = 28.9 \,\, \left(\frac{V_0}{100 \text{\ km/s}}\right)^{-3/4} \left(\frac{\rho(0)}{10^{-24} \text{\ g/cm}^3}\right)^{1/4}\ \text{eV}\,.
\end{equation}

This limit is similar to the phase-density constraints \cite{Boyanovsky2008Constraints,Tremaine1979Dynamical,Madsen1990Phase,Madsen1991Generalized} but slightly less by a numerical factor.
Eq.~(\ref{eq:mLowerBEC}) is founded on the BEC existence criterion and is quite distinct from the principle that the maximum phase-space density decreases through the Universe evolution.

The phase-density constraints are generally considered to obtain a lower bound for WDM candidates. It is derived when applied to ultra-faint dwarf galaxies, which are detected as satellites around the Milky Way. This kind of galaxies are very faint, very dense and considered to be the smallest DM dominated structures with the central density $\rho\sim 10^{-23}\ \text{g/cm}^3$ and the velocity dispersion $v\sim 5 \ \text{km/s}$ \cite{Simon2011COMPLETE,Willman2012GALAXY}. The corresponding phase-density constraints give the particle mass range above a few keV.

The phase-density approach is directly applied for fermions because the Fermi-Dirac distribution has the maximum value \cite{Tremaine1979Dynamical}. Contrary to fermions, there is no fixed maximum in the Bose--Einstein distribution, and this makes a barrier to directly employ the method for bosons, particularly when BEC is present \cite{Boyanovsky2008Constraints,Madsen1990Phase,Madsen1991Generalized}. Modifications of the phase-density approach regarding the bosons imply homogeneous involvement of the condensed and noncondensed particles. Taking into account the upper bound from (\ref{eq:mUpper}) one obtains for the keV bosons $g^*_\text{d}>300$, which indicates that the DM bosons decouple well above the electroweak scales \cite{Boyanovsky2008Constraints}.

In our model relation (\ref{eq:mLowerBEC}) when directly applied to the compact dwarfs yields the mass value of at least 0.5 keV. This estimation, however, supposes that the Bose stars are not a significant or at least comparable fraction in the satellite galaxies.
A more reasonable approach appears to be that constraint~(\ref{eq:mLowerBEC}) is accepted for larger dwarfs and LSB galaxies while the ultra-faint dwarfs are to be considered as pure Bose star clusters with a negligible contribution of the nondegenerate fraction. 
To put it another way, the structure of supermassive globular star clusters is assumed to reproduce in the ultra-faint dwarf galaxies by the Bose stars. In this case taking the central density $\rho\sim 10^{-24}\ \text{g/cm}^3$ and the velocity dispersion $V_0\sim 25 \ \text{km/s}$ one finds the lower mass bound is about 80~eV. The number of degrees of freedom consistent with this mass, $g^*_\text{d}>25$, is obtained from (\ref{eq:mUpper}). It corresponds to DM decoupling just after the QCD phase transition.

\section{Confronting with rotation curve data}
\label{sec:RotationCurves}

To test the validity of our model we confront it with rotation curves data. Needless to say, the majority of the observed rotation curves contain contributions from various galaxy components apart from the DM, such as disk, gas, and bulge. In this case the fitting procedure involves additional free parameters associated with the baryonic sector.
The prime interest of our analysis is to verify whether the DM model with the dominant noncondensed component provides observable properties of the DM halos. At this point in our research we present a preliminary analysis for the rotation curves of LSB galaxies, which are overwhelmingly dominated by DM \cite{Blok1997dark}. In this respect the LSB galaxies provide an opportunity for testing predictions regarding DM models. Keeping in mind only a minor contribution of the baryonic component to the overall LSB galaxy mass we ignore it in our simulations.

We selected 22 LSB galaxies from the data sample considered in \cite{McGaugh2001High} with a good $H\alpha$ rotation-curve quality and observations close to the galactic center ($\lesssim$ 15 kpc).

First, we calculate the rotation curves for the pure noncondensed DM model represented by Eq.~(\ref{eq:Bose-gas}). In this case the model contains three independent parameters. One of them, the particle mass  $m$,  is a model parameter while the other two qualify as chemical parameters, i.e. describing only a DM distribution in the halo. For these free parameters, one can take, say, the central density $\rho_0$ and the scale radius $R$ as in heuristic profiles below.

We also compare our results with the two-parametric DM profiles such as the IS, pseudo-isothermal (PIS), and Burkert (Bur) profiles. These profiles represent a spatial mass distribution of DM in the halos although fail to involve information about the nature of DM. 
Each of the profiles has a fixed shape, which is scaled by two parameters;  those tend to be the central density $\rho_0$ and the scale radius $R$.
Nevertheless, they can fit good to a large number of the rotation curves.


\begin{table}[!ht]
	\begin{tabular}{|c|c||c|c|c|c|c|c|c|c|c|c|}
		\toprule
		No & Galaxy &
		$ \;\;\;\; R \;\;\;\;$&	
		$ \;\;\;\; V_0 \;\;\;\;$ &
		$ \;\;\;\; \nu \;\;\;\; $  &
		$ \;\;\;\; m \;\;\;\;$&
		\multicolumn{6}{c|}{$\chi^2_{\text{r}}$} \\ \cline{7-12}
		& & (kpc) & 	(km/s) &  & (eV) &
		min & 86 eV & IS & PIS & Bur & NFW
		\\  \cline{3-8}
		& & \multicolumn{6}{c|}{Noncondensed DM bosons} &
		&  &  &  \\
		\hline
		(1)&(2)&(3)&(4)&(5)&(6)&(7)&(8)&(9)&(10)&(11)&(12)\\ \botrule%
		1 & \text{F563-1} & 1.1 & 73.5 & 2.61 & 57 & 0.12 & 0.22 & 0.24 & 0.09 & 0.19 & 0.09 \\
		2 & \text{F568-3} & 1.6 & 64.5 & 0 & --- & 0.28 & 0.30 & 0.28 & 0.52 & 0.44 & 2.25 \\
		3 & \textbf{F571-8} & 1.2 & 95.0 & 2.61 & 51 & 1.24 & \textbf{3.01} & 3.22 & 1.52 & 2.19 & 1.50 \\
		4 & \text{F579-V1} & 0.4 & 74.2 & 2.6 & 89 & 0.14 & 0.16 & 0.45 & 0.03 & 0.43 & 0.21 \\
		5 & \text{F583-1} & 1.4 & 57.1 & 1.03 & 66 & 0.003 & 0.01 & 0.02 & 0.04 & 0.003 & 0.74 \\
		6 & \text{F583-4} & 0.6 & 44.1 & 2.61 & 86 & 0.43 & 0.43 & 0.83 & 0.33 & 0.66 & 0.32 \\
		7 & \text{F730-V1} & 0.8 & 93.6 & 2.61 & 62 & 0.19 & 0.72 & 0.95 & 0.10 & 0.60 & 1.00 \\
		8 & \text{UGC-4115} & 1.1 & 59.2 & 2.46 & 61 &\ 0.002\ \ &\ 0.003\ \ &\ 0.004\ \ &\ 0.004\ \ &\ 0.002\ \ &\ 0.80\ \ \\
		9 & \text{UGC-5750} & 3.5 & 73.8 & 2.61 & 31 & 1.23 & 1.26 & 1.26 & 1.25 & 1.24 & 2.21 \\
		10 & \text{UGC-11454} & 1.1 & 101.6 & 2.14 & 54 & 0.67 & 1.26 & 1.41 & 0.42 & 0.97 & 3.34 \\
		11 & \text{UGC-11557} & 3.7 & 102.6 & 2.21 & 29 & 0.05 & 0.06 & 0.06 & 0.05 & 0.05 & 1.62 \\
		12 & \text{UGC-11583} & 0.4 & 23.9 & 0 & --- & 0.08 & 0.23 & 0.08 & 0.1 & 0.11 & 0.73 \\
		13 & \text{UGC-11616} & 0.8 & 88.8 & 1.9 & 67 & 0.24 & 0.34 & 0.47 & 0.14 & 0.34 & 1.27 \\
		14 &\textbf{UGC-11648} & 1.1 & 88.6 & 2.61 & 53 &\ 4.05\ & \textbf{6.43} & 6.72 & 3.81 & 5.50 & 0.97 \\
		15 & \text{UGC-11819} & 1.6 & 100.7 & 0 & --- & 0.17 & 0.17 & 0.17 & 0.30 & 0.18 & 1.35 \\
		16 & \text{ESO-0840411} & 3.9 & 70.3 & 2.61 & 30 & 0.03 & 0.08 & 0.08 & 0.07 & 0.05 & 1.87 \\
		17 & \text{ESO-1870510} & 0.4 & 30.0 & 2.61 & 111 & 0.02 & 0.18 & 0.05 & 0.03 & 0.03 & 0.06 \\
		18 & \text{ESO-2060140} & 0.6 & 77.9 & 2.61 & 74 & 0.08 & 0.17 & 0.30 & 0.11 & 0.19 & 0.42 \\
		19 & \text{ESO-3020120} & 1.2 & 54.3 & 0.06 & 152 & 0.002 & 0.003 & 0.002 & 0.04 & 0.006 & 0.33 \\
		20 & \text{ESO-3050090} & 1.3 & 47.8 & 2.61 & 58 & 0.03 & 0.06 & 0.06 & 0.05 & 0.04 & 0.24 \\
		21 & \text{ESO-4250180} & 2.2 & 102.5 & 2.61 & 37 & 0.08 & 0.13 & 0.13 & 0.09 & 0.10 & 0.01 \\
		22 &\ \text{ESO-4880049}\ \ & 1.0 & 69.4 & 2.61 & 59 & 0.01 & 0.07 & 0.09  & 0.02 & 0.04 & 0.17 \\
		
		\botrule
	\end{tabular}
	\caption{The best-fit parameters for the 22 LSB galaxies.}\label{tab:fitting nc}
\end{table}


The IS profile is an asymptotic version for the pure noncondensed DM model when $\nu\rightarrow 0$ and represent a self-gravitating ideal gas as it is described by Eq.~(\ref{eq:ISM}). In fitting procedures its approximation
\begin{equation}\label{eq:PISprofile}
\rho_\text{PIS} = \rho_0\left(1+\frac{r^2}{R^2}\right)^{-1}\,,
\end{equation}
known as PIS density profile, is more widely used within the phenomenological approach.

Yet another heuristic DM profile is given by
\begin{equation}
\rho_\text{Bur} = \rho_0\left(1+\frac{r}{R}\right)^{-1}\left(1+\frac{r^2}{R^2}\right)^{-1}\,,
\end{equation}
proposed by Burkert \cite{Burkert1995Structure}, which is cored close to center and provides Navarro--Frenk--White (NFW)-like behavior at large $r$.

The NFW profile
\begin{equation}
\rho_\text{NFW} = \rho_0\left(\frac{r}{R}\right)^{-1}\left(1+\frac{r}{R}\right)^{-2}\,
\end{equation}
is an approximation for DM halo distribution produced by $N$-body simulation for collisionless DM \cite{Navarro1995Structure}. It has a divergent central density so $\rho_0$ is merely a density parameter. The NFW profile is well suited to massive galaxies and galaxy clusters and is worse consistent with observations of LSB galaxies \cite{McGaugh2001High}. Nevertheless it is used as universal profile to describe the DM distribution in a wide range of galactic masses.

The numerical dimensionless velocity profile is calculated as the integral
\begin{equation}\label{eq:circV}
u^2(\xi)=\frac{1}{\xi}\int\limits_0^\xi x(s)s^2{\,{\rm{d}}} s\,
\end{equation}
where $x(s)$ evolves according to Eq.~(\ref{eq:Bose-gas}). It is related to the circular velocity of the bosons as
\begin{equation}\label{eq:Vcalc}
V(r) = V_0 u(r/R)\,,
\end{equation}
where $V_0$ is defined by (\ref{eq:dispV0}).

The simulated velocity (\ref{eq:Vcalc}) is fitted to the observed rotation curves minimizing the residual sum
\begin{equation}
\chi^2 = \sum\limits_{i=1}^N \frac{\left[V(r_i)-V_\text{obs}(r_i)\right]^2}{\sigma^2(r_i)}\,,
\end{equation}
where $V_\text{obs}$ and $\sigma$ give the observed rotation velocities and corresponding observational uncertainties, and $N$ is the number of observations.

In practical calculations $R$, $V_0$ and $\nu$ are more appropriate fitting parameters. Values of  $R$  and $V_0$ are positive and fitted without an upper limit while $\nu$ is limited by the condensation value. The particle mass can be obtained through (\ref{eq:inhR}) and (\ref{eq:Vcalc}) as
\begin{equation}\label{eq:massfit}
m=\left(\sqrt{\frac\pi 2}\frac{\hbar^3}{G\nu R^2 V_0}\right)^{1/4}\,.
\end{equation}
For the two-parametric profiles only the first two parameters are used.


\begin{figure}[hpt!]
	\begin{minipage}[h]{0.03\linewidth}
		\rotatebox{90}{rotation velocity (km/s)}
	\end{minipage}	
	\hfill
	\begin{minipage}[h]{0.96\linewidth}
		\begin{minipage}[h]{0.47\linewidth}
			\includegraphics[width=1\linewidth]{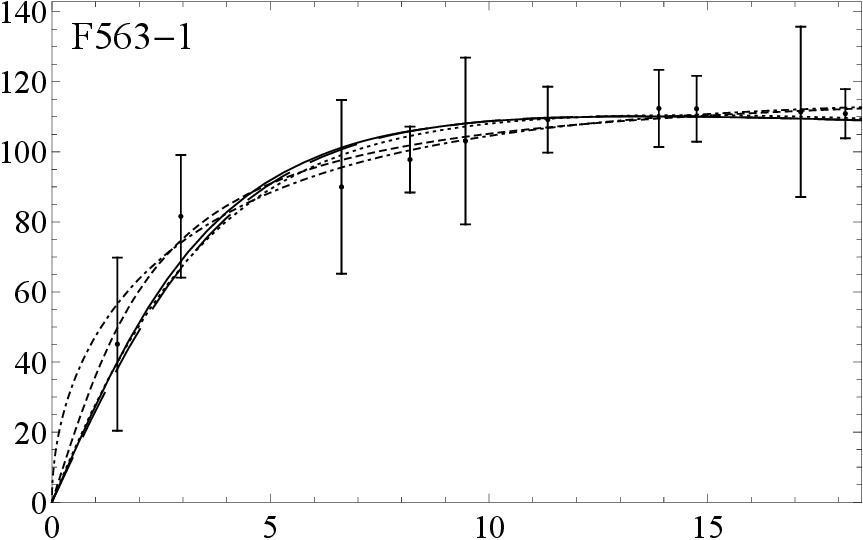}
		\end{minipage}
		\hfill
		\begin{minipage}[h]{0.47\linewidth}
			\includegraphics[width=1\linewidth]{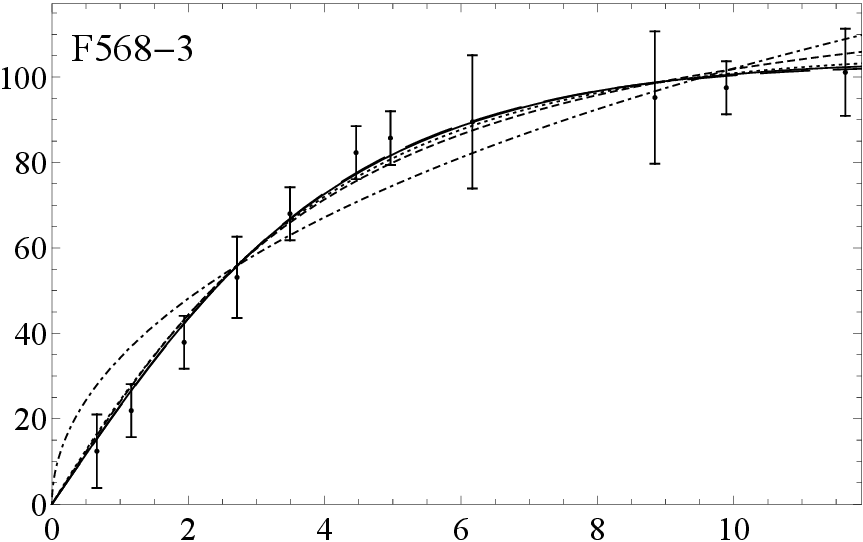}
		\end{minipage}
		\vfill
		\vspace{0.03\linewidth}
		\begin{minipage}[h]{0.47\linewidth}
			\includegraphics[width=1\linewidth]{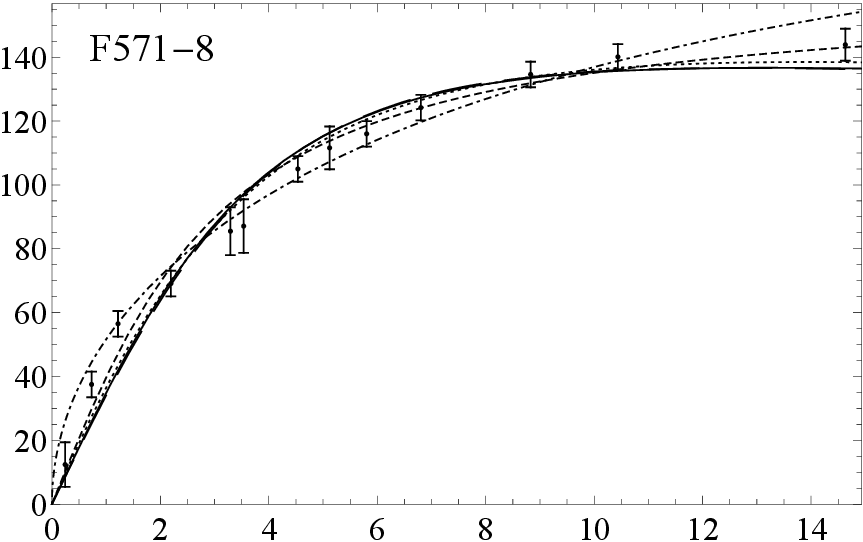}
		\end{minipage}
		\hfill
		\begin{minipage}[h]{0.47\linewidth}
			\includegraphics[width=1\linewidth]{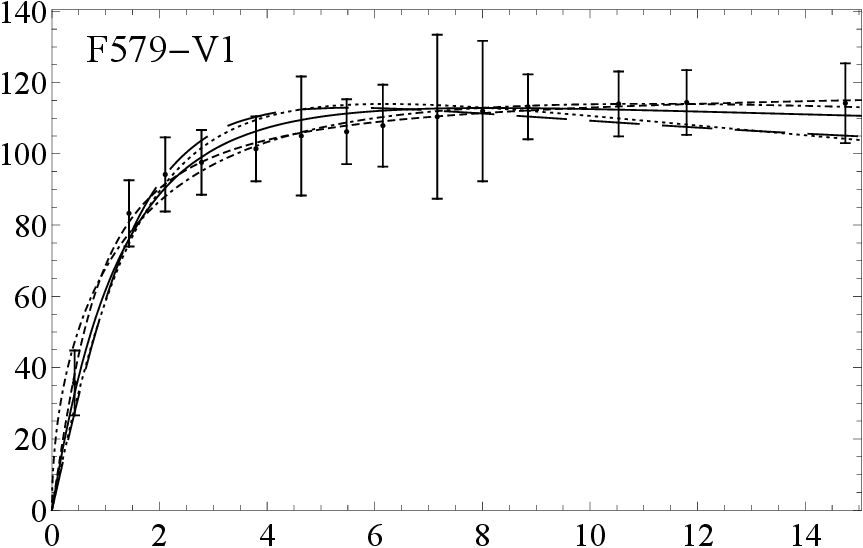}
		\end{minipage}
		\begin{minipage}[h]{0.47\linewidth}
			\vspace{0.07\linewidth}
			\includegraphics[width=1\linewidth]{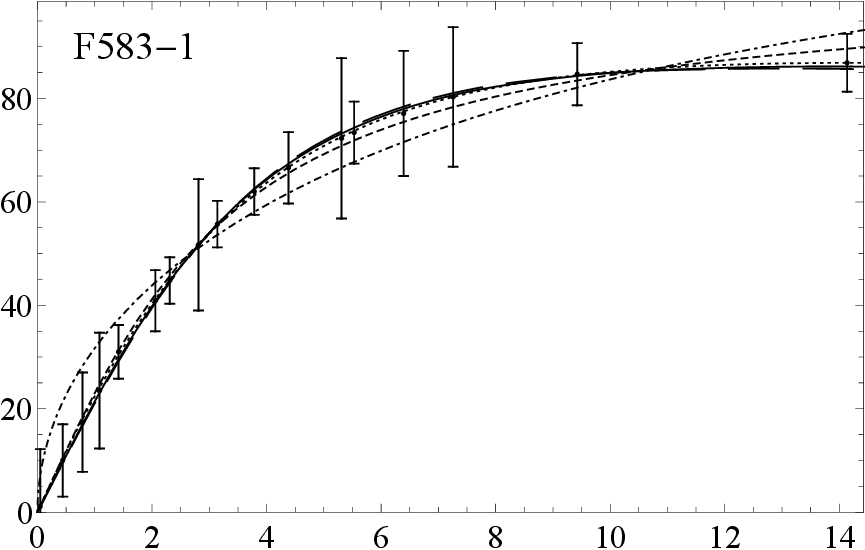}
		\end{minipage}
		\hfill
		\begin{minipage}[h]{0.47\linewidth}
			\vspace{0.07\linewidth}
			\includegraphics[width=1\linewidth]{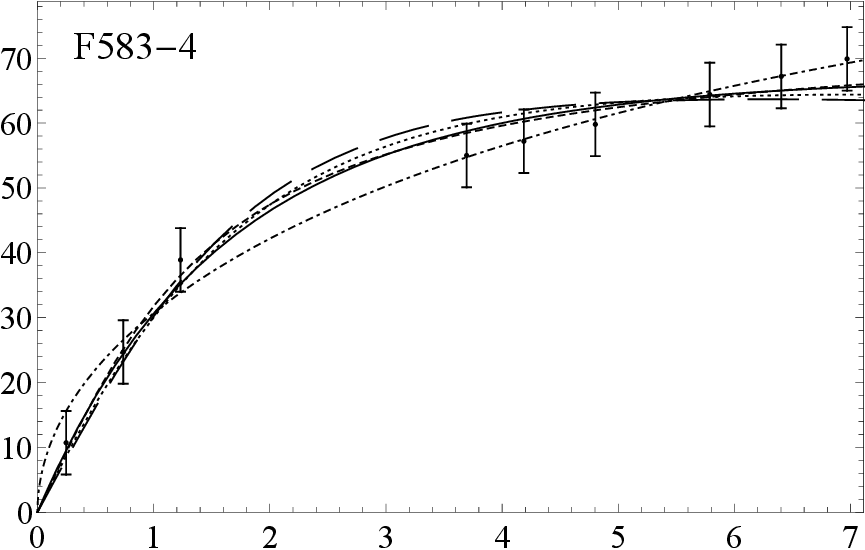}
		\end{minipage}
		\vfill
		\vspace{0.03\linewidth}
		\begin{minipage}[h]{0.47\linewidth}
			\includegraphics[width=1\linewidth]{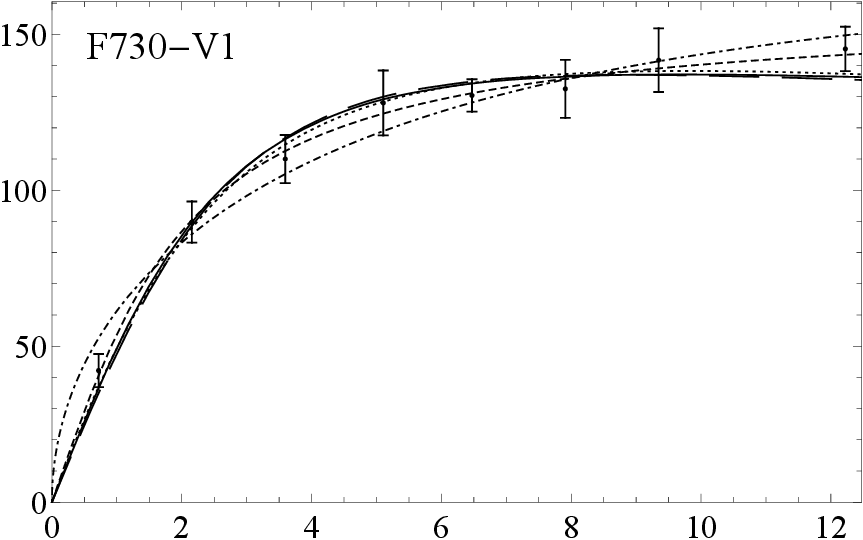}
		\end{minipage}
		\hfill
		\begin{minipage}[h]{0.47\linewidth}
			\includegraphics[width=1\linewidth]{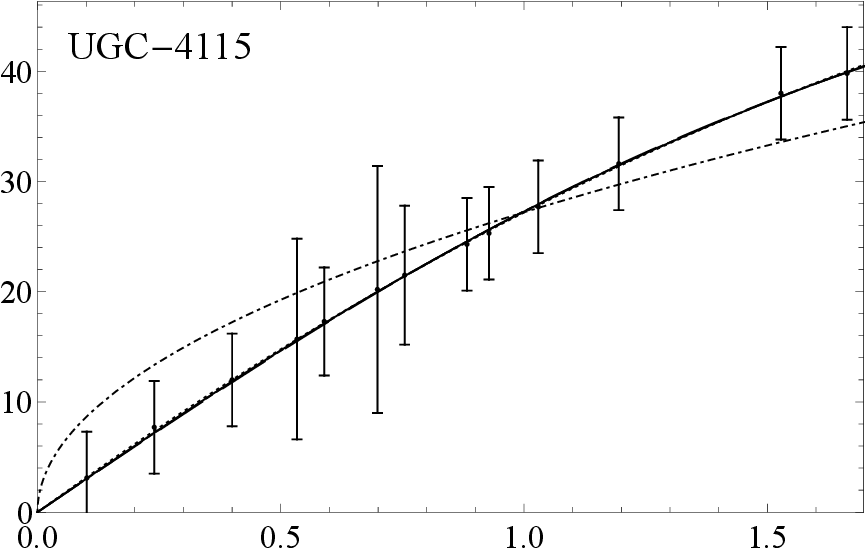}
		\end{minipage}
	\vfill
	\vspace{0.02\linewidth}
		{radial distance (kpc)}
			\vspace{0.02\linewidth}
	\end{minipage}
	\caption[vcplots]{The best-fit rotation curves for the IS (long dashed line), PIS (dashed line), Burkert (dotted line), and NFW (dot-dashed line) profiles, and the pure noncondensed DM model at $m = 86$~eV (solid line).}
	\label{fig:vcplots1}
\end{figure}
\addtocounter{figure}{-1}

\begin{figure}[hpt!]
	\begin{minipage}[h]{0.03\linewidth}
		\rotatebox{90}{ rotation velocity (km/s)}
	\end{minipage}	
	\hfill
	\begin{minipage}[h]{0.96\linewidth}
		\begin{minipage}[h]{0.47\linewidth}
			\includegraphics[width=1\linewidth]{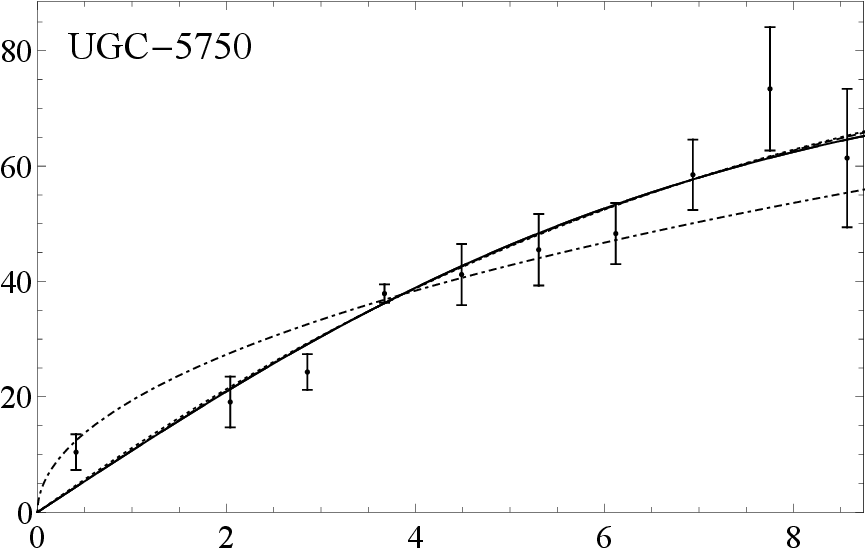}
		\end{minipage}
		\hfill
		\begin{minipage}[h]{0.47\linewidth}
			\includegraphics[width=1\linewidth]{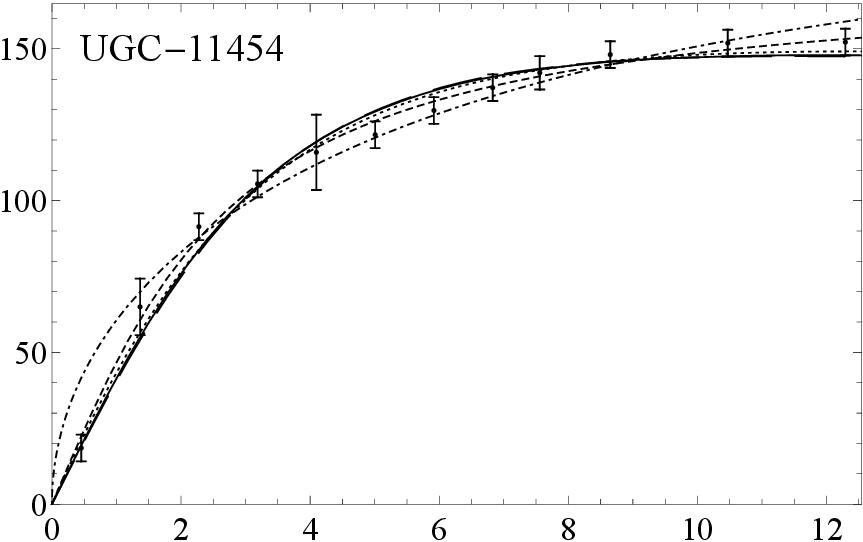}
		\end{minipage}
		\vfill
		\vspace{0.03\linewidth}
		\begin{minipage}[h]{0.47\linewidth}
			\includegraphics[width=1\linewidth]{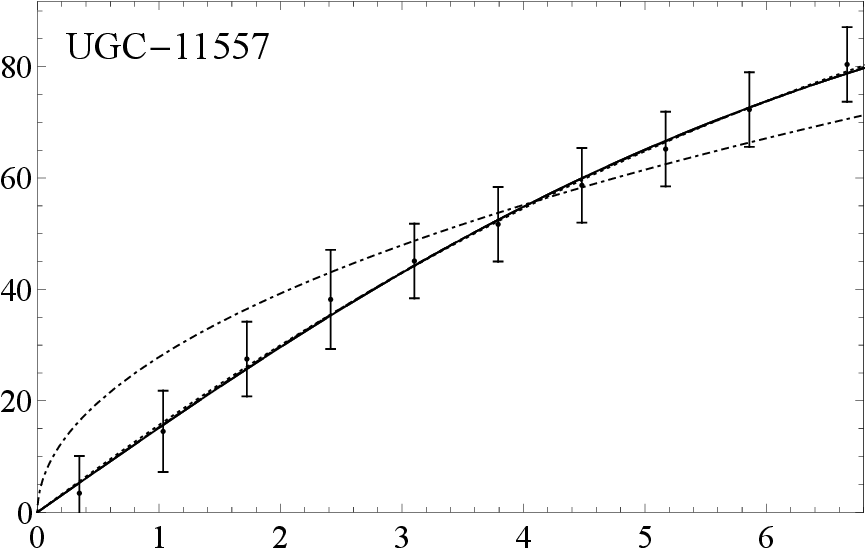}
		\end{minipage}
		\hfill
		\begin{minipage}[h]{0.47\linewidth}
			\includegraphics[width=1\linewidth]{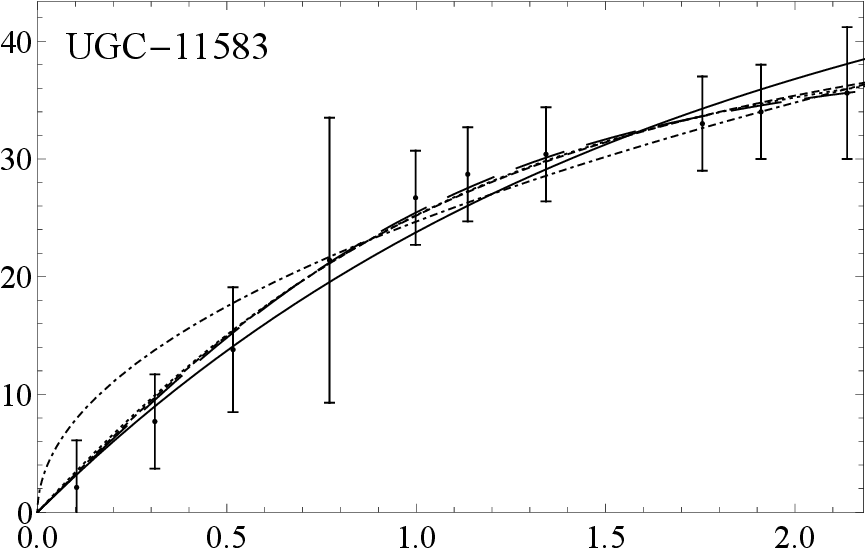}
		\end{minipage}
		\begin{minipage}[h]{0.47\linewidth}
			\vspace{0.07\linewidth}
			\includegraphics[width=1\linewidth]{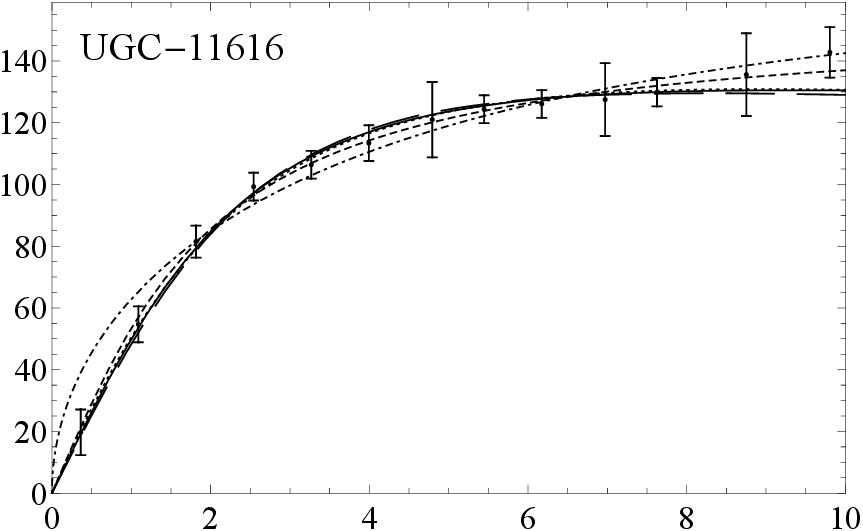}
		\end{minipage}
		\hfill
		\begin{minipage}[h]{0.47\linewidth}
			\vspace{0.07\linewidth}
			\includegraphics[width=1\linewidth]{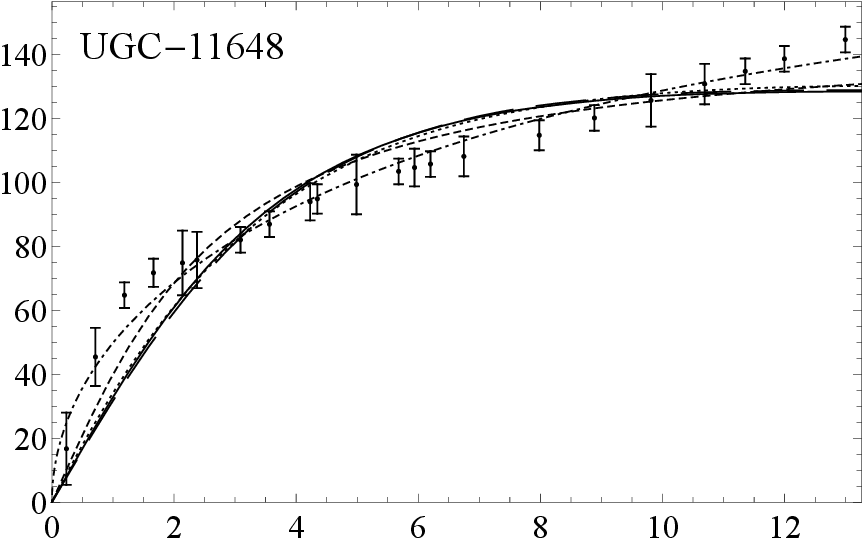}
		\end{minipage}
		\vfill
		\vspace{0.03\linewidth}
		\begin{minipage}[h]{0.47\linewidth}
			\includegraphics[width=1\linewidth]{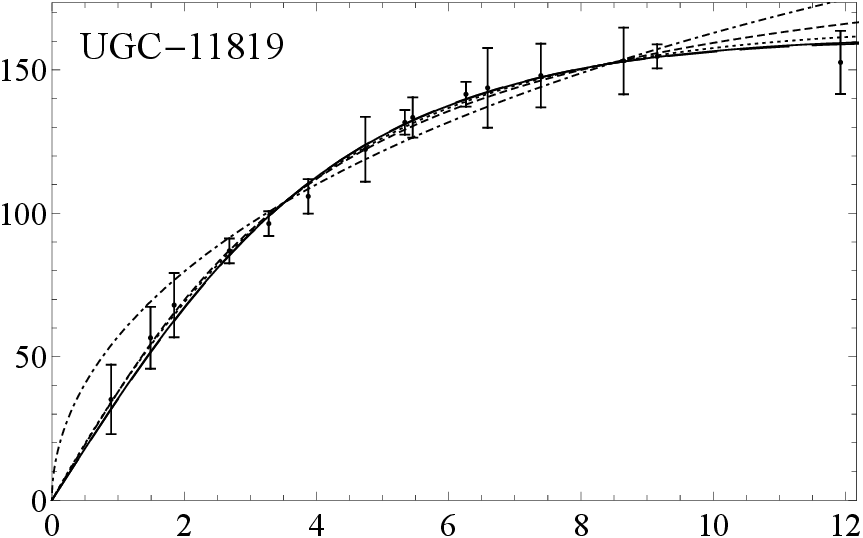}
		\end{minipage}
		\hfill
		\begin{minipage}[h]{0.47\linewidth}
			\includegraphics[width=1\linewidth]{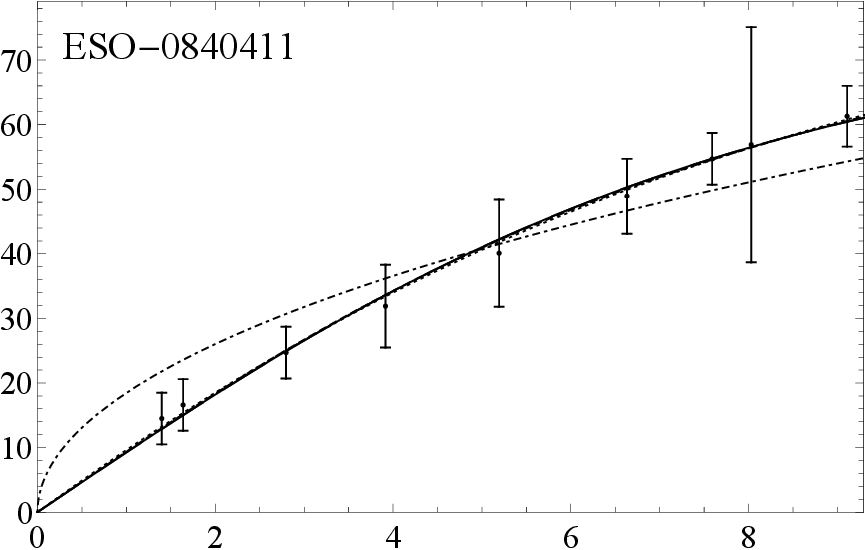}
		\end{minipage}
		\vfill
		\vspace{0.02\linewidth}
		{radial distance (kpc)}
		\vspace{0.02\linewidth}
	\end{minipage}
	\caption[vcplots]{Continuation}
	\label{fig:vcplots2}
\end{figure}
\addtocounter{figure}{-1}

\begin{figure}[hpt!]
	\begin{minipage}[h]{0.03\linewidth}
		\rotatebox{90}{rotation velocity (km/s)} 
	\end{minipage}	
	\hfill
	\begin{minipage}[h]{0.96\linewidth}
		\begin{minipage}[h]{0.47\linewidth}
			\includegraphics[width=1\linewidth]{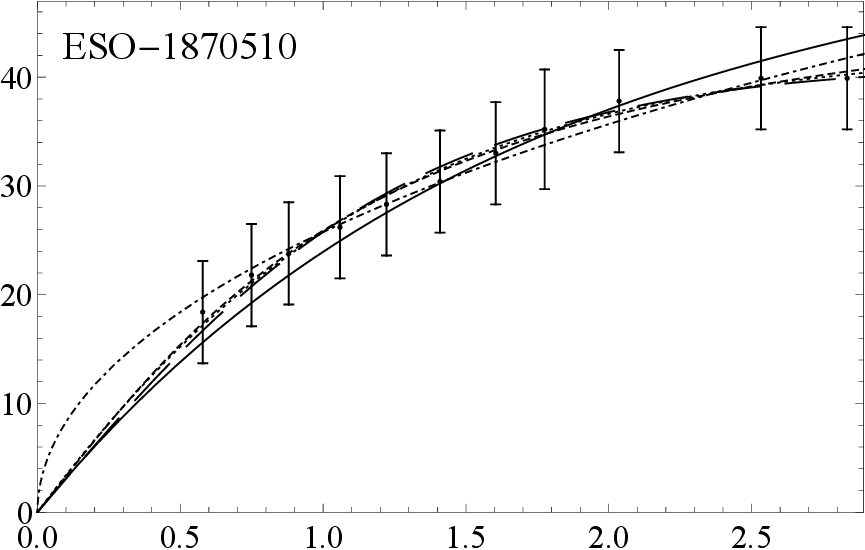}
		\end{minipage}
		\hfill
		\begin{minipage}[h]{0.47\linewidth}
			\includegraphics[width=1\linewidth]{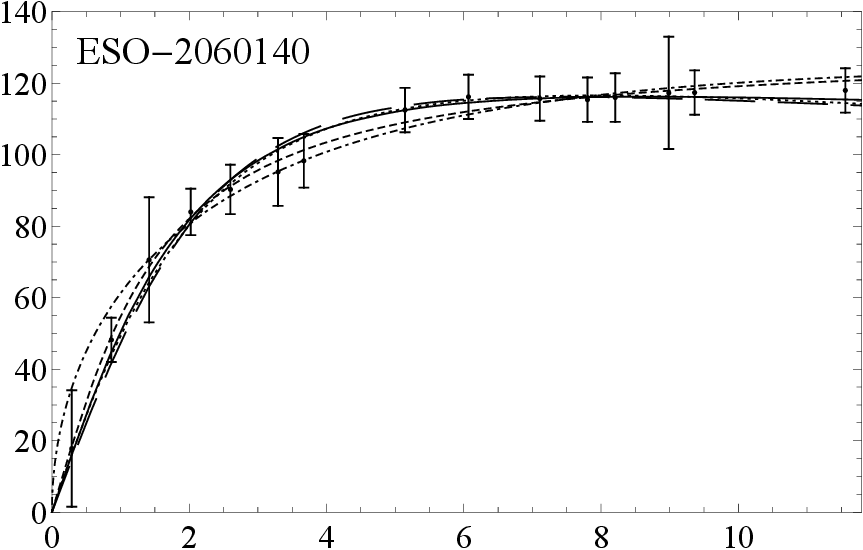}
		\end{minipage}
		\vfill
		\vspace{0.03\linewidth}
		\begin{minipage}[h]{0.47\linewidth}
			\includegraphics[width=1\linewidth]{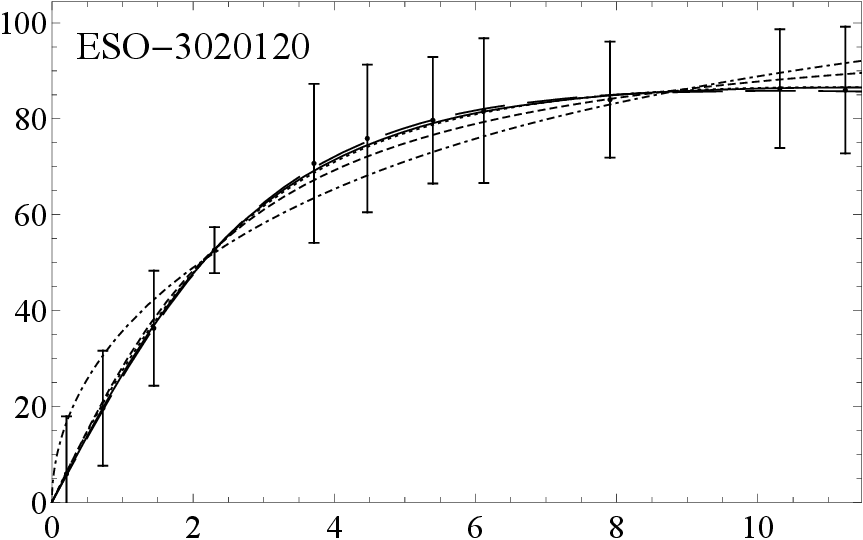}
		\end{minipage}
		\hfill
		\begin{minipage}[h]{0.47\linewidth}
			\includegraphics[width=1\linewidth]{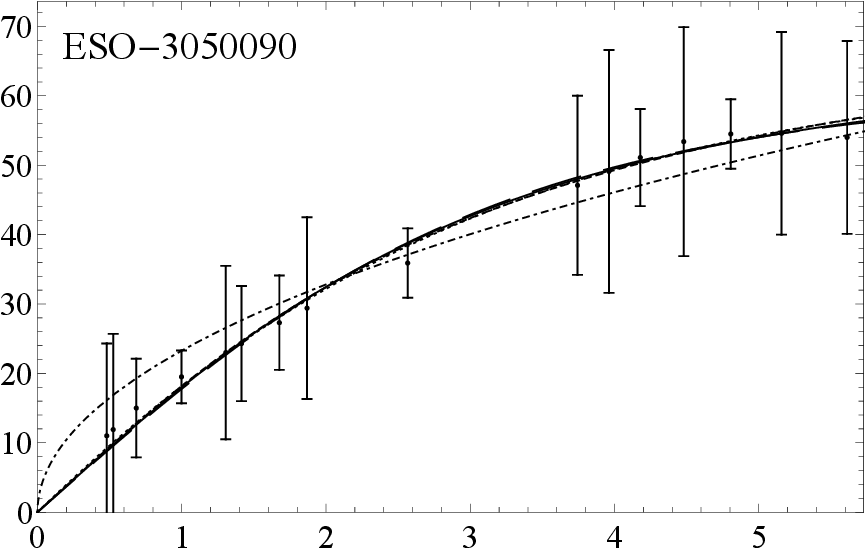}
		\end{minipage}
		\begin{minipage}[h]{0.47\linewidth}
			\vspace{0.07\linewidth}
			\includegraphics[width=1\linewidth]{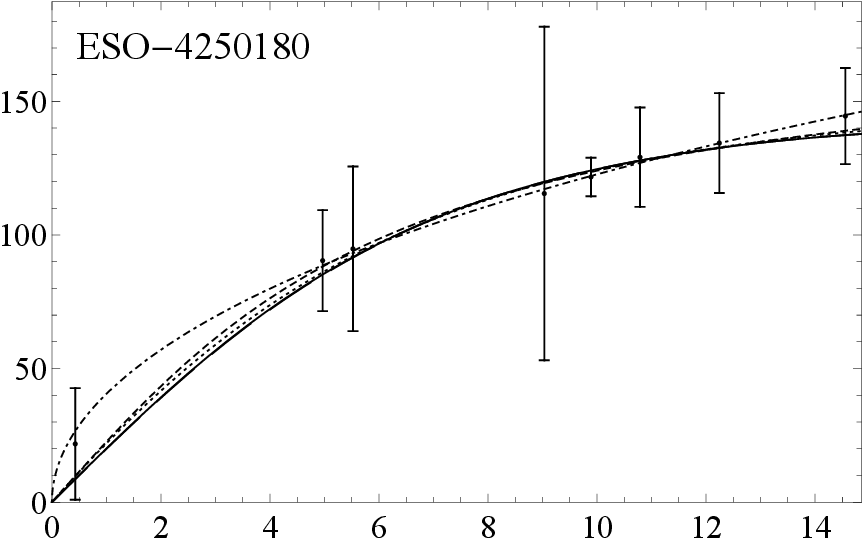}
		\end{minipage}
		\hfill
		\begin{minipage}[h]{0.47\linewidth}
			\vspace{0.07\linewidth}
			\includegraphics[width=1\linewidth]{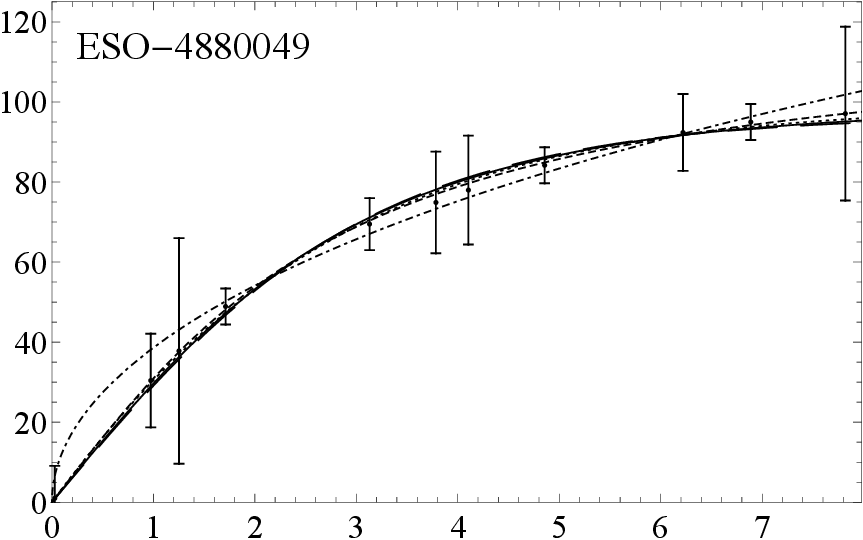}
		\end{minipage}
		
		\vspace{0.02\linewidth}		
		{radial distance (kpc)}
		\vspace{0.02\linewidth}	
	\end{minipage}

	\caption[vcplots]{Continuation}
	\label{fig:vcplots3}
\end{figure}


Table~\ref{tab:fitting nc} shows the best-fit parameters for each of the considered galaxies for the pure noncondensed DM model.
The fitting parameters are shown in columns (3)--(5). The corresponding particle mass is given in column (6). Three galaxies give the best-fit $\nu=0$, which corresponds to IS model. In these cases the particle mass remains indeterminate. The other galaxies give $\nu>0$, and in most cases $\nu$ turns out to be equal to its critical value $\nu=\zeta(3/2)$. For these galaxies the values of the mass fall within the range of 29~eV$\leqslant m \leqslant$152~eV.

Columns (7)--(12) present values of the reduced residual sums $\chi^2_\text{r}=\chi^2/N_\text{dof}$ for each of the estimated DM models, where $N_\text{dof}$ is the corresponding number degrees of freedom. The comparison between the observed and the best-fit rotation curves according to the considered DM models for each galaxy from the sample is also shown in Fig.~\ref{fig:vcplots1}. It is seen that the pure noncondensed DM model agrees well with the observational
curves for LSB galaxies to the same extent as the cored empirical DM profiles and significantly better than the NFW profile.
This is the evidence in favor of that the noncondensed boson DM model correctly describes the galaxy structures.

The majority of the galaxies from the sample demonstrate a good fit for the particle mass range over 50 eV. It is  verified by $\chi^2$ test and is clearly read with the rule of thumb $\chi^2_\text{r}\lesssim 1$.  Only two galaxies, F571-8 and UGC-11648  (marked out by the bold font in the Table~\ref{tab:fitting nc}), give an unsatisfactory result with the completely noncondensed DM.

We also study $\chi^2$ for the noncondensed DM bosons depending on the particle mass. For the large masses results of rotation curves fitting for the nondegenerate bosons is indistinguishable from the IS model.  This is the expectable result since the condition $\nu\ll 1$ is obeyed in this case according to (\ref{eq:nu}) and (\ref{eq:deBroglie}), and Eq.~(\ref{eq:Bose-gas}) transforms into Eq.~(\ref{eq:ISM}). When the particle mass decreases, $\chi^2$ runs into its minimum  (its reduced values are given in Table~\ref{tab:fitting nc}) and grows very steeply thereafter. Such behavior comes from the fact that at the minimum $\chi^2$ the parameter $\nu$ has its critical value or reaches it very fast when the mass decreases according to (\ref{eq:massfit}). In this case two of three parameters are fixed, and we can manipulate the only parameter, say $R$, to fit the rotation curve, so that the residual uncertainties remain rather large.

\begin{figure}
	\centering
	\includegraphics[width=0.7\linewidth]{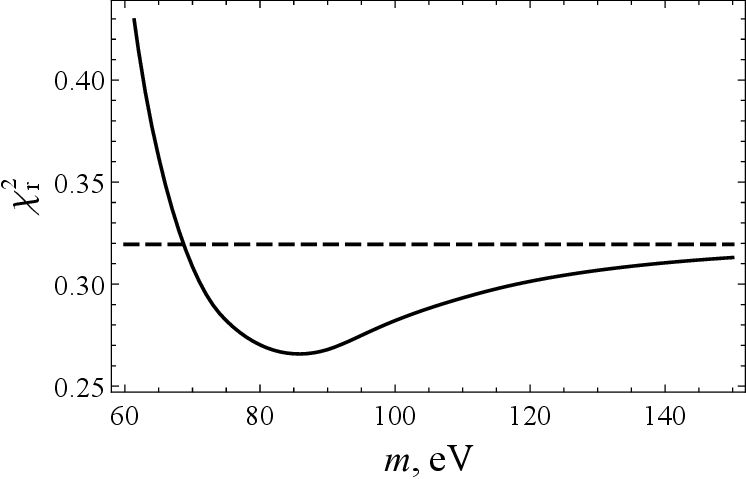}
	\caption{The total residual sum $\chi_\text{r}^2$ depending on the particle mass for the pure noncondensed DM halos. The sum is taken over the shortened sample including the 20 galaxies with a good fit. The horizontal dashed line represents the equivalent value of $\chi_\text{r}^2$ for the IS model.}
	\label{fig:X2mass}
\end{figure}

This dependence is reproduced in Fig.~\ref{fig:X2mass} for the total reduced residual sum $\chi^2_\text{r}$ taken over the entire sample excluding the two galaxies with a poor fit. The best-fit particle mass obtained for the shortened sample including the 20 galaxies is 86 eV. Corresponding values of $\chi^2_\text{r}$ for each galaxy are presented in column (8) of Table~\ref{tab:fitting nc} and the rotation curves are shown in Fig.~\ref{fig:vcplots1} .

\begin{table}
	\begin{tabular}{|c|c||c|c|c|c|c|c|c|c|c|}
		\toprule
	Galaxy &	
	$ \;\;\;\; R \;\;\;\;$&	
	$ \;\;\;\; V_0 \;\;\;\;$ &
	$ \;\;\;\; \nu \;\;\;\; $  &
	$ \;\;\;\; \alpha \;\;\;\;$&
	\multicolumn{2}{c|}{$\chi^2_{\text{r}}$} \\ \cline{6-7}
	 & (kpc) & 	(km/s) &  &  &\ {\small two-component}\ \ &\  {\small noncondensed only}\ \
				\\ \toprule
	\text{F571-8} & 1.5 & 93.2 & 0.19 & 12.8 & 0.92 & 3.01 \\
	\ \text{UGC-11648}\ \  & 2.3 & 101.7 & 0.08 & 5.9 & 0.46 & 6.43 \\
		\botrule
	\end{tabular}
	\caption{The best-fit parameters at the boson mass $m=86$~eV for the two LSB galaxies taking into account the Bose star component.}\label{tab:fitting 2comp}
\end{table}

The observational data for the 20 galaxies are consistent with the parameters obtained for the pure noncondensed DM model and dispenses with the need for further adjustments. To improve the results for the two galaxies with the poor fit we employ Eqs.~(\ref{eq:2components}) describing the DM halo with both the condensed and noncondensed fractions. The two-component version of the fitting model includes an additional parameter describing a Bose stars distribution in relation to the nondegenerate fraction. We set the exact equality between the spatial scaling factor $\gamma$ and the central densities ratio $\alpha$ throughout the fitting procedure to keep it within the scope of the constraint  specified by (\ref{eq:RatioCentrDens}). The particle mass remains fixed at the value $m=86$ eV.  The circular velocity is found by integral (\ref{eq:circV}) with $x+x_\text{bs}$ instead of $x$. The obtained parameters are presented in Table~\ref{tab:fitting 2comp}. It is clear that the observational data for these two galaxies are well described by the two-component model, which provides $\chi_\text{r}^2 <1$.

\begin{figure}[hpt!]
	\begin{minipage}[h]{0.03\linewidth}
		\rotatebox{90}{\qquad\ rotation velocity (km/s)}
	\end{minipage}	
	\hfill
	\begin{minipage}[h]{0.96\linewidth}
		\begin{minipage}[h]{0.47\linewidth}
			\includegraphics[width=1\linewidth]{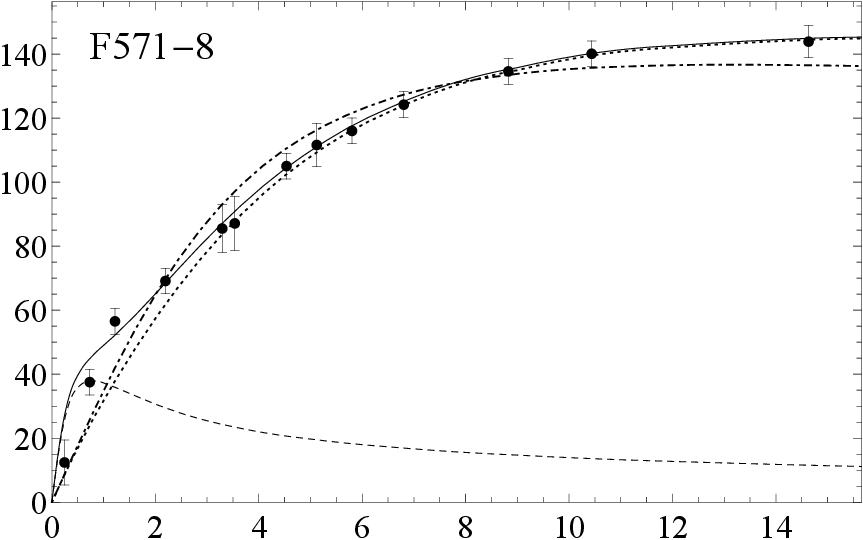}
		\end{minipage}
		\hfill
		\begin{minipage}[h]{0.47\linewidth}
			\includegraphics[width=1\linewidth]{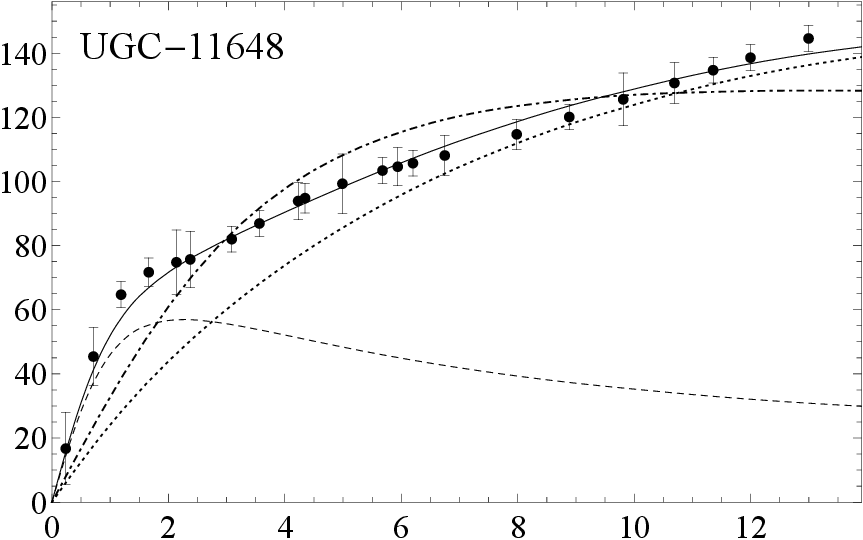}
		\end{minipage}
		\vfill
		\vspace{0.02\linewidth}
		{radial distance (kpc)}
		\vspace{0.02\linewidth}	
	\end{minipage}
	\caption{The best-fit rotation curves for the two-component boson DM. The solid line shows the total circular velocity, the dashed and dotted lines correspond to contributions of the condensed (Bose stars) and noncondensed fraction respectively. The rotation curve for the completely noncondensed DM (dot-dashed line) is given for comparison. }
\label{fig:vplots2}
\end{figure}

The corresponding rotation curves with the contribution produced by each component are shown in Fig.~\ref{fig:vplots2}. It is seen that the model curves get better fit by virtue of the denser Bose star distribution in the halo center as it was discussed in Section~\ref{sec:BS}.

\section{Conclusions}

We have examined the model of the DM halos, which consist of the condensed and noncondensed bosons. This approach provides the galaxy structures in agreement with observations when the particle mass is in the range above about 50~eV.

In the scope of this model a considerable fraction of the DM bosons is noncondensed in the halo. However,  the condensed particles are significant in galaxy structure formation. The condensed, zero momentum bosons form the smallest DM entities in this model, compact BECs, also named Bose stars. These objects are considerably less than galaxies in size and mass for both the self-interacting and non-interacting particles. Although the self-interaction is unimportant on the galaxy scale it supports against gravity when very dense clumps transform into BECs. For the non-interacting bosons the quantum pressure has the same part at even smaller scales.

The nondegenerate bosons close to the zero momentum state are localized near the galaxy center.  In this region the DM is dense enough to provide the particles to be close to the critical point and  follow the Bose--Einstein distribution. For the particles with mass within the range above 50~eV this regime is supported inside the radius between one to a few kpc. This extent corresponds to dwarf galaxies and determines a DM halo core for larger LSB galaxies.

Outside the core the DM becomes diluted, so it is described by the Maxwell--Boltzmann distribution as usual. The diluted regime is also typical for large galaxies. In this case the bosons reproduce a picture corresponding to the IS model of CDM halos.

Contrasting the theoretical rotation curves with those obtained from observations, we found that the completely noncondensed halo approximation fits well a number of galaxies and is highly competitive with the widely used heuristic core density profiles such as the PIS and Burkert models. The steeper growth of the circular velocities at small distances and attendant irregularities in the rising part of the rotation curves can be described within the two-component approach, which includes the noncondensed bosons and the clusters of BECs.

A fraction consisting of the Bose stars brings a wide variety in the galaxy structures. Regions with a large population of the Bose stars provide a considerable contribution of the condensed bosons to the total halo mass. It seems reasonable that BECs are mostly assembled in miniclusters, which accumulate inside the core 
resulting in the steeper growth of the rotation velocities against the pure noncondensed halo.

Outside the core, BECs are expected to appear in small amounts. However, it is not improbable that the BEC miniclusters can be more evenly distributed over the halo and contribute far from the galaxy center to the density profile.  The large clusters are likely detected as ultra-faint dwarf galaxies, which are observed among Milky Way satellites.

The particle mass range under discussion is vastly below the WIMP mass, which enables us to classify the particles as light bosons. On the other hand, the particles considerably exceed axions and ultralight bosons in mass and fall into the mass range corresponding to WDM. Nevertheless, the BEC fraction is to be considered as CDM due to zero particle momenta. Thus, a CDM/WDM mixture naturally arises for this kind of bosons. These particles are acceptable to be thermally produced and have to decouple at the QCD transition epoch to provide the observed DM abundance today.

Speaking about applications of this model in cosmological $N$-body simulations one expects a similar result as for WIMPs at the galaxy scales while BEC minicluster formation should be taken into account for the  subgalactic scales.

\acknowledgments

The work was supported by Russian Foundation for Basic Research (Project No 20-52-05009), and partially by the Program of Competitive Growth of Kazan Federal University.

\newpage

\bibliography{lit2}

\begin{thebibliography}{83}%
\makeatletter
\providecommand \@ifxundefined [1]{%
 \@ifx{#1\undefined}
}%
\providecommand \@ifnum [1]{%
 \ifnum #1\expandafter \@firstoftwo
 \else \expandafter \@secondoftwo
 \fi
}%
\providecommand \@ifx [1]{%
 \ifx #1\expandafter \@firstoftwo
 \else \expandafter \@secondoftwo
 \fi
}%
\providecommand \natexlab [1]{#1}%
\providecommand \enquote  [1]{``#1''}%
\providecommand \bibnamefont  [1]{#1}%
\providecommand \bibfnamefont [1]{#1}%
\providecommand \citenamefont [1]{#1}%
\providecommand \href@noop [0]{\@secondoftwo}%
\providecommand \href [0]{\begingroup \@sanitize@url \@href}%
\providecommand \@href[1]{\@@startlink{#1}\@@href}%
\providecommand \@@href[1]{\endgroup#1\@@endlink}%
\providecommand \@sanitize@url [0]{\catcode `\\12\catcode `\$12\catcode
  `\&12\catcode `\#12\catcode `\^12\catcode `\_12\catcode `\%12\relax}%
\providecommand \@@startlink[1]{}%
\providecommand \@@endlink[0]{}%
\providecommand \url  [0]{\begingroup\@sanitize@url \@url }%
\providecommand \@url [1]{\endgroup\@href {#1}{\urlprefix }}%
\providecommand \urlprefix  [0]{URL }%
\providecommand \Eprint [0]{\href }%
\providecommand \doibase [0]{http://dx.doi.org/}%
\providecommand \selectlanguage [0]{\@gobble}%
\providecommand \bibinfo  [0]{\@secondoftwo}%
\providecommand \bibfield  [0]{\@secondoftwo}%
\providecommand \translation [1]{[#1]}%
\providecommand \BibitemOpen [0]{}%
\providecommand \bibitemStop [0]{}%
\providecommand \bibitemNoStop [0]{.\EOS\space}%
\providecommand \EOS [0]{\spacefactor3000\relax}%
\providecommand \BibitemShut  [1]{\csname bibitem#1\endcsname}%
\let\auto@bib@innerbib\@empty
\bibitem [{\citenamefont {Bertone}\ \emph {et~al.}(2005)\citenamefont
  {Bertone}, \citenamefont {Hooper},\ and\ \citenamefont
  {Silk}}]{Bertone2005Particle}%
  \BibitemOpen
  \bibfield  {author} {\bibinfo {author} {\bibfnamefont {G.}~\bibnamefont
  {Bertone}}, \bibinfo {author} {\bibfnamefont {D.}~\bibnamefont {Hooper}}, \
  and\ \bibinfo {author} {\bibfnamefont {J.}~\bibnamefont {Silk}},\ }\href
  {\doibase 10.1016/j.physrep.2004.08.031} {\bibfield  {journal} {\bibinfo
  {journal} {\emph{Phys. Rep.}}\ }\textbf {\bibinfo {volume} {405}},\ \bibinfo
  {pages} {279} (\bibinfo {year} {2005})}\BibitemShut {NoStop}%
\bibitem [{\citenamefont {Mukhanov}(2005)}]{Mukhanov2005Physical}%
  \BibitemOpen
  \bibfield  {author} {\bibinfo {author} {\bibfnamefont {V.}~\bibnamefont
  {Mukhanov}},\ }\href
  {https://www.ebook.de/de/product/4016569/p_j_steinhardt_viatcheslav_mukhanov_physical_foundations_of_cosmology.html}
  {\emph {\bibinfo {title} {Physical foundations of cosmology}}}\ (\bibinfo
  {publisher} {Cambridge University Press},\ \bibinfo {year}
  {2005})\BibitemShut {NoStop}%
\bibitem [{\citenamefont {Navarro}(1996)}]{Navarro1996structure}%
  \BibitemOpen
  \bibfield  {author} {\bibinfo {author} {\bibfnamefont {J.~F.}\ \bibnamefont
  {Navarro}},\ }in\ \href {\doibase 10.1017/s0074180900232452} {\emph {\bibinfo
  {booktitle} {Symposium-international astronomical union}}},\ Vol.\ \bibinfo
  {volume} {171}\ (\bibinfo {organization} {Cambridge University Press},\
  \bibinfo {year} {1996})\ pp.\ \bibinfo {pages} {255--258}\BibitemShut
  {NoStop}%
\bibitem [{\citenamefont {Bullock}\ and\ \citenamefont
  {Boylan-Kolchin}(2017)}]{Bullock2017Small}%
  \BibitemOpen
  \bibfield  {author} {\bibinfo {author} {\bibfnamefont {J.~S.}\ \bibnamefont
  {Bullock}}\ and\ \bibinfo {author} {\bibfnamefont {M.}~\bibnamefont
  {Boylan-Kolchin}},\ }\href {\doibase 10.1146/annurev-astro-091916-055313}
  {\bibfield  {journal} {\bibinfo  {journal} {\emph{Annu. Rev. Astron.
  Astrophys.}}\ }\textbf {\bibinfo {volume} {55}},\ \bibinfo {pages} {343}
  (\bibinfo {year} {2017})}\BibitemShut {NoStop}%
\bibitem [{\citenamefont {de~Blok}(2010)}]{Blok2010Core}%
  \BibitemOpen
  \bibfield  {author} {\bibinfo {author} {\bibfnamefont {W.~J.~G.}\
  \bibnamefont {de~Blok}},\ }\href {\doibase 10.1155/2010/789293} {\bibfield
  {journal} {\bibinfo  {journal} {\emph{Advances in Astronomy}}\ }\textbf
  {\bibinfo {volume} {2010}},\ \bibinfo {pages} {1} (\bibinfo {year}
  {2010})}\BibitemShut {NoStop}%
\bibitem [{\citenamefont {Strigari}\ \emph {et~al.}(2007)\citenamefont
  {Strigari} \emph {et~al.}}]{Strigari2007Redefining}%
  \BibitemOpen
  \bibfield  {author} {\bibinfo {author} {\bibfnamefont {L.~E.}\ \bibnamefont
  {Strigari}} \emph {et~al.},\ }\href {\doibase 10.1086/521914} {\bibfield
  {journal} {\bibinfo  {journal} {\emph{Astrophys. J.}}\ }\textbf {\bibinfo
  {volume} {669}},\ \bibinfo {pages} {676} (\bibinfo {year}
  {2007})}\BibitemShut {NoStop}%
\bibitem [{\citenamefont {Su{\'a}rez}\ \emph {et~al.}(2014)\citenamefont
  {Su{\'a}rez}, \citenamefont {Robles},\ and\ \citenamefont
  {Matos}}]{Suarez2014review}%
  \BibitemOpen
  \bibfield  {author} {\bibinfo {author} {\bibfnamefont {A.}~\bibnamefont
  {Su{\'a}rez}}, \bibinfo {author} {\bibfnamefont {V.~H.}\ \bibnamefont
  {Robles}}, \ and\ \bibinfo {author} {\bibfnamefont {T.}~\bibnamefont
  {Matos}},\ }in\ \href {\doibase 10.1007/978-3-319-02063-1} {\emph {\bibinfo
  {booktitle} {Accelerated cosmic expansion}}},\ \bibinfo {editor} {edited by\
  \bibinfo {editor} {\bibfnamefont {C.}~\bibnamefont {Moreno~Gonz{\'a}lez}},
  \bibinfo {editor} {\bibfnamefont {J.~E.}\ \bibnamefont {Madriz~Aguilar}}, \
  and\ \bibinfo {editor} {\bibfnamefont {L.~M.}\ \bibnamefont
  {Reyes~Barrera}}}\ (\bibinfo  {publisher} {Springer International
  Publishing},\ \bibinfo {year} {2014})\ pp.\ \bibinfo {pages}
  {107--142}\BibitemShut {NoStop}%
\bibitem [{\citenamefont {Hu}\ \emph {et~al.}(2000)\citenamefont {Hu},
  \citenamefont {Barkana},\ and\ \citenamefont {Gruzinov}}]{Hu2000Fuzzy}%
  \BibitemOpen
  \bibfield  {author} {\bibinfo {author} {\bibfnamefont {W.}~\bibnamefont
  {Hu}}, \bibinfo {author} {\bibfnamefont {R.}~\bibnamefont {Barkana}}, \ and\
  \bibinfo {author} {\bibfnamefont {A.}~\bibnamefont {Gruzinov}},\ }\href
  {\doibase 10.1103/physrevlett.85.1158} {\bibfield  {journal} {\bibinfo
  {journal} {\emph{Phys. Rev. Lett.}}\ }\textbf {\bibinfo {volume} {85}},\
  \bibinfo {pages} {1158} (\bibinfo {year} {2000})}\BibitemShut {NoStop}%
\bibitem [{\citenamefont {Lee}\ and\ \citenamefont {Koh}(1996)}]{LeeBS}%
  \BibitemOpen
  \bibfield  {author} {\bibinfo {author} {\bibfnamefont {J.}~\bibnamefont
  {Lee}}\ and\ \bibinfo {author} {\bibfnamefont {I.}~\bibnamefont {Koh}},\
  }\href {\doibase 10.1103/PhysRevD.53.2236} {\bibfield  {journal} {\bibinfo
  {journal} {\emph{Phys. Rev. D}}\ }\textbf {\bibinfo {volume} {53}},\ \bibinfo
  {pages} {2236} (\bibinfo {year} {1996})}\BibitemShut {NoStop}%
\bibitem [{\citenamefont {Boehmer}\ and\ \citenamefont
  {Harko}(2007)}]{Boehmer2007Can}%
  \BibitemOpen
  \bibfield  {author} {\bibinfo {author} {\bibfnamefont {C.}~\bibnamefont
  {Boehmer}}\ and\ \bibinfo {author} {\bibfnamefont {T.}~\bibnamefont
  {Harko}},\ }\href {\doibase 10.1088/1475-7516/2007/06/025} {\bibfield
  {journal} {\bibinfo  {journal} {\emph{J. Cosmol. Astropart. Phys.}}\ }\textbf
  {\bibinfo {volume} {2007}},\ \bibinfo {pages} {025} (\bibinfo {year}
  {2007})}\BibitemShut {NoStop}%
\bibitem [{\citenamefont {Slepian}\ and\ \citenamefont
  {Goodman}(2012)}]{Slepian2012Ruling}%
  \BibitemOpen
  \bibfield  {author} {\bibinfo {author} {\bibfnamefont {Z.}~\bibnamefont
  {Slepian}}\ and\ \bibinfo {author} {\bibfnamefont {J.}~\bibnamefont
  {Goodman}},\ }\href {\doibase 10.1111/j.1365-2966.2012.21901.x} {\bibfield
  {journal} {\bibinfo  {journal} {\emph{Mon. Not. R. Astron. Soc.}}\ }\textbf
  {\bibinfo {volume} {427}},\ \bibinfo {pages} {839} (\bibinfo {year}
  {2012})}\BibitemShut {NoStop}%
\bibitem [{\citenamefont {Seidel}\ and\ \citenamefont
  {Suen}(1990)}]{Seidel1990Dynamical}%
  \BibitemOpen
  \bibfield  {author} {\bibinfo {author} {\bibfnamefont {E.}~\bibnamefont
  {Seidel}}\ and\ \bibinfo {author} {\bibfnamefont {W.}~\bibnamefont {Suen}},\
  }\href {\doibase 10.1103/PhysRevD.42.384} {\bibfield  {journal} {\bibinfo
  {journal} {\emph{Phys. Rev. D}}\ }\textbf {\bibinfo {volume} {42}},\ \bibinfo
  {pages} {384} (\bibinfo {year} {1990})}\BibitemShut {NoStop}%
\bibitem [{\citenamefont {Sin}(1994)}]{Sin1994Late}%
  \BibitemOpen
  \bibfield  {author} {\bibinfo {author} {\bibfnamefont {S.}~\bibnamefont
  {Sin}},\ }\href {\doibase 10.1103/physrevd.50.3650} {\bibfield  {journal}
  {\bibinfo  {journal} {\emph{Phys. Rev. D}}\ }\textbf {\bibinfo {volume}
  {50}},\ \bibinfo {pages} {3650} (\bibinfo {year} {1994})}\BibitemShut
  {NoStop}%
\bibitem [{\citenamefont {Schive}\ \emph {et~al.}(2014)\citenamefont {Schive},
  \citenamefont {Chiueh},\ and\ \citenamefont {Broadhurst}}]{Schive2014Cosmic}%
  \BibitemOpen
  \bibfield  {author} {\bibinfo {author} {\bibfnamefont {H.}~\bibnamefont
  {Schive}}, \bibinfo {author} {\bibfnamefont {T.}~\bibnamefont {Chiueh}}, \
  and\ \bibinfo {author} {\bibfnamefont {T.}~\bibnamefont {Broadhurst}},\
  }\href {\doibase 10.1038/nphys2996} {\bibfield  {journal} {\bibinfo
  {journal} {\emph{Nat. Phys.}}\ }\textbf {\bibinfo {volume} {10}},\ \bibinfo
  {pages} {496} (\bibinfo {year} {2014})}\BibitemShut {NoStop}%
\bibitem [{\citenamefont {Du}\ \emph {et~al.}(2016)\citenamefont {Du},
  \citenamefont {Behrens},\ and\ \citenamefont
  {Niemeyer}}]{Du2016Substructure}%
  \BibitemOpen
  \bibfield  {author} {\bibinfo {author} {\bibfnamefont {X.}~\bibnamefont
  {Du}}, \bibinfo {author} {\bibfnamefont {C.}~\bibnamefont {Behrens}}, \ and\
  \bibinfo {author} {\bibfnamefont {J.~C.}\ \bibnamefont {Niemeyer}},\ }\href
  {\doibase 10.1093/mnras/stw2724} {\bibfield  {journal} {\bibinfo  {journal}
  {\emph{Mon. Not. R. Astron. Soc.}}\ }\textbf {\bibinfo {volume} {465}},\
  \bibinfo {pages} {941} (\bibinfo {year} {2016})}\BibitemShut {NoStop}%
\bibitem [{\citenamefont {Press}\ \emph {et~al.}(1990)\citenamefont {Press},
  \citenamefont {Ryden},\ and\ \citenamefont {Spergel}}]{Press1990Single}%
  \BibitemOpen
  \bibfield  {author} {\bibinfo {author} {\bibfnamefont {W.~H.}\ \bibnamefont
  {Press}}, \bibinfo {author} {\bibfnamefont {B.~S.}\ \bibnamefont {Ryden}}, \
  and\ \bibinfo {author} {\bibfnamefont {D.~N.}\ \bibnamefont {Spergel}},\
  }\href {\doibase 10.1103/PhysRevLett.64.1084} {\bibfield  {journal} {\bibinfo
   {journal} {\emph{Phys. Rev. Lett.}}\ }\textbf {\bibinfo {volume} {64}},\
  \bibinfo {pages} {1084} (\bibinfo {year} {1990})}\BibitemShut {NoStop}%
\bibitem [{\citenamefont {Lee}\ and\ \citenamefont
  {Lim}(2010)}]{Lee2010Minimum}%
  \BibitemOpen
  \bibfield  {author} {\bibinfo {author} {\bibfnamefont {J.}~\bibnamefont
  {Lee}}\ and\ \bibinfo {author} {\bibfnamefont {S.}~\bibnamefont {Lim}},\
  }\href {\doibase 10.1088/1475-7516/2010/01/007} {\bibfield  {journal}
  {\bibinfo  {journal} {\emph{J. Cosmol. Astropart. Phys.}}\ }\textbf {\bibinfo
  {volume} {2010}},\ \bibinfo {pages} {007} (\bibinfo {year}
  {2010})}\BibitemShut {NoStop}%
\bibitem [{\citenamefont {Safarzadeh}\ and\ \citenamefont
  {Spergel}(2020)}]{Safarzadeh2020Ultra}%
  \BibitemOpen
  \bibfield  {author} {\bibinfo {author} {\bibfnamefont {M.}~\bibnamefont
  {Safarzadeh}}\ and\ \bibinfo {author} {\bibfnamefont {D.~N.}\ \bibnamefont
  {Spergel}},\ }\href {\doibase 10.3847/1538-4357/ab7db2} {\bibfield  {journal}
  {\bibinfo  {journal} {\emph{Astrophys. J.}}\ }\textbf {\bibinfo {volume}
  {893}},\ \bibinfo {pages} {21} (\bibinfo {year} {2020})}\BibitemShut
  {NoStop}%
\bibitem [{\citenamefont {Sikivie}\ \emph {et~al.}(2007)\citenamefont
  {Sikivie}, \citenamefont {Tanner},\ and\ \citenamefont {van
  Bibber}}]{throuWall}%
  \BibitemOpen
  \bibfield  {author} {\bibinfo {author} {\bibfnamefont {P.}~\bibnamefont
  {Sikivie}}, \bibinfo {author} {\bibfnamefont {D.~B.}\ \bibnamefont {Tanner}},
  \ and\ \bibinfo {author} {\bibfnamefont {K.}~\bibnamefont {van Bibber}},\
  }\href {http://dx.doi.org/10.1103/PhysRevLett.98.172002} {\bibfield
  {journal} {\bibinfo  {journal} {\emph{Phys. Rev. Lett.}}\ }\textbf {\bibinfo
  {volume} {98}} (\bibinfo {year} {2007})}\BibitemShut {NoStop}%
\bibitem [{\citenamefont {Ehret}\ \emph {et~al.}(2010)\citenamefont {Ehret}
  \emph {et~al.}}]{ALPS}%
  \BibitemOpen
  \bibfield  {author} {\bibinfo {author} {\bibfnamefont {K.}~\bibnamefont
  {Ehret}} \emph {et~al.},\ }\href {\doibase
  https://doi.org/10.1016/j.physletb.2010.04.066} {\bibfield  {journal}
  {\bibinfo  {journal} {\emph{Phys. Lett. B}}\ }\textbf {\bibinfo {volume}
  {689}},\ \bibinfo {pages} {149} (\bibinfo {year} {2010})}\BibitemShut
  {NoStop}%
\bibitem [{\citenamefont {Arik}\ \emph {et~al.}(2014)\citenamefont {Arik} \emph
  {et~al.}}]{CAST}%
  \BibitemOpen
  \bibfield  {author} {\bibinfo {author} {\bibfnamefont {M.}~\bibnamefont
  {Arik}} \emph {et~al.} (\bibinfo {collaboration} {{CAST} Collaboration}),\
  }\href {\doibase 10.1103/PhysRevLett.112.091302} {\bibfield  {journal}
  {\bibinfo  {journal} {\emph{Phys. Rev. Lett.}}\ }\textbf {\bibinfo {volume}
  {112}},\ \bibinfo {pages} {091302} (\bibinfo {year} {2014})}\BibitemShut
  {NoStop}%
\bibitem [{\citenamefont {Caldwell}\ \emph {et~al.}(2017)\citenamefont
  {Caldwell} \emph {et~al.}}]{Caldwell2017Dielectric}%
  \BibitemOpen
  \bibfield  {author} {\bibinfo {author} {\bibfnamefont {A.}~\bibnamefont
  {Caldwell}} \emph {et~al.},\ }\href
  {https://doi.org/10.1103/PhysRevLett.118.091801} {\bibfield  {journal}
  {\bibinfo  {journal} {\emph{Phys. Rev. Lett.}}\ }\textbf {\bibinfo {volume}
  {118}} (\bibinfo {year} {2017})}\BibitemShut {NoStop}%
\bibitem [{\citenamefont {Du}\ \emph {et~al.}(2018)\citenamefont {Du} \emph
  {et~al.}}]{ADMX}%
  \BibitemOpen
  \bibfield  {author} {\bibinfo {author} {\bibfnamefont {N.}~\bibnamefont {Du}}
  \emph {et~al.} (\bibinfo {collaboration} {{ADMX} Collaboration}),\ }\href
  {\doibase 10.1103/PhysRevLett.120.151301} {\bibfield  {journal} {\bibinfo
  {journal} {\emph{Phys. Rev. Lett.}}\ }\textbf {\bibinfo {volume} {120}},\
  \bibinfo {pages} {151301} (\bibinfo {year} {2018})}\BibitemShut {NoStop}%
\bibitem [{\citenamefont {Peccei}\ and\ \citenamefont {Quinn}(1977)}]{PQ}%
  \BibitemOpen
  \bibfield  {author} {\bibinfo {author} {\bibfnamefont {R.~D.}\ \bibnamefont
  {Peccei}}\ and\ \bibinfo {author} {\bibfnamefont {H.~R.}\ \bibnamefont
  {Quinn}},\ }\href {\doibase 10.1103/PhysRevLett.38.1440} {\bibfield
  {journal} {\bibinfo  {journal} {\emph{Phys. Rev. Lett.}}\ }\textbf {\bibinfo
  {volume} {38}},\ \bibinfo {pages} {1440} (\bibinfo {year}
  {1977})}\BibitemShut {NoStop}%
\bibitem [{\citenamefont {Duffy}\ and\ \citenamefont {van
  Bibber}(2009)}]{Duffy2009Axions}%
  \BibitemOpen
  \bibfield  {author} {\bibinfo {author} {\bibfnamefont {L.~D.}\ \bibnamefont
  {Duffy}}\ and\ \bibinfo {author} {\bibfnamefont {K.}~\bibnamefont {van
  Bibber}},\ }\href {\doibase 10.1088/1367-2630/11/10/105008} {\bibfield
  {journal} {\bibinfo  {journal} {\emph{New J. Phys.}}\ }\textbf {\bibinfo
  {volume} {11}},\ \bibinfo {pages} {105008} (\bibinfo {year}
  {2009})}\BibitemShut {NoStop}%
\bibitem [{\citenamefont {Erken}\ \emph {et~al.}(2012)\citenamefont {Erken},
  \citenamefont {Sikivie}, \citenamefont {Tam},\ and\ \citenamefont
  {Yang}}]{Erken2012Axion}%
  \BibitemOpen
  \bibfield  {author} {\bibinfo {author} {\bibfnamefont {O.}~\bibnamefont
  {Erken}}, \bibinfo {author} {\bibfnamefont {P.}~\bibnamefont {Sikivie}},
  \bibinfo {author} {\bibfnamefont {H.}~\bibnamefont {Tam}}, \ and\ \bibinfo
  {author} {\bibfnamefont {Q.}~\bibnamefont {Yang}},\ }\href {\doibase
  10.1103/physrevlett.108.061304} {\bibfield  {journal} {\bibinfo  {journal}
  {\emph{Phys. Rev. Lett.}}\ }\textbf {\bibinfo {volume} {108}},\ \bibinfo
  {pages} {061304} (\bibinfo {year} {2012})}\BibitemShut {NoStop}%
\bibitem [{\citenamefont {Conlon}(2006)}]{Conlon2006The}%
  \BibitemOpen
  \bibfield  {author} {\bibinfo {author} {\bibfnamefont {J.~P.}\ \bibnamefont
  {Conlon}},\ }\href {\doibase 10.1088/1126-6708/2006/05/078} {\bibfield
  {journal} {\bibinfo  {journal} {\emph{J. High Energ. Phys.}}\ }\textbf
  {\bibinfo {volume} {2006}},\ \bibinfo {pages} {078} (\bibinfo {year}
  {2006})}\BibitemShut {NoStop}%
\bibitem [{\citenamefont {Svrcek}\ and\ \citenamefont
  {Witten}(2006)}]{Svrcek2006Axions}%
  \BibitemOpen
  \bibfield  {author} {\bibinfo {author} {\bibfnamefont {P.}~\bibnamefont
  {Svrcek}}\ and\ \bibinfo {author} {\bibfnamefont {E.}~\bibnamefont
  {Witten}},\ }\href {\doibase 10.1088/1126-6708/2006/06/051} {\bibfield
  {journal} {\bibinfo  {journal} {\emph{J. High Energ. Phys.}}\ }\textbf
  {\bibinfo {volume} {2006}},\ \bibinfo {pages} {051} (\bibinfo {year}
  {2006})}\BibitemShut {NoStop}%
\bibitem [{\citenamefont {Arvanitaki}\ \emph {et~al.}(2010)\citenamefont
  {Arvanitaki} \emph {et~al.}}]{Arvanitaki2010String}%
  \BibitemOpen
  \bibfield  {author} {\bibinfo {author} {\bibfnamefont {A.}~\bibnamefont
  {Arvanitaki}} \emph {et~al.},\ }\href {\doibase 10.1103/PhysRevD.81.123530}
  {\bibfield  {journal} {\bibinfo  {journal} {\emph{Phys. Rev. D}}\ }\textbf
  {\bibinfo {volume} {81}},\ \bibinfo {pages} {123530} (\bibinfo {year}
  {2010})}\BibitemShut {NoStop}%
\bibitem [{\citenamefont {Graham}\ \emph {et~al.}(2015)\citenamefont {Graham},
  \citenamefont {Kaplan},\ and\ \citenamefont
  {Rajendran}}]{Graham2015Cosmological}%
  \BibitemOpen
  \bibfield  {author} {\bibinfo {author} {\bibfnamefont {P.~W.}\ \bibnamefont
  {Graham}}, \bibinfo {author} {\bibfnamefont {D.~E.}\ \bibnamefont {Kaplan}},
  \ and\ \bibinfo {author} {\bibfnamefont {S.}~\bibnamefont {Rajendran}},\
  }\href {\doibase 10.1103/PhysRevLett.115.221801} {\bibfield  {journal}
  {\bibinfo  {journal} {\emph{Phys. Rev. Lett.}}\ }\textbf {\bibinfo {volume}
  {115}},\ \bibinfo {pages} {221801} (\bibinfo {year} {2015})}\BibitemShut
  {NoStop}%
\bibitem [{\citenamefont {Espinosa}\ \emph {et~al.}(2015)\citenamefont
  {Espinosa} \emph {et~al.}}]{Espinosa2015Cosmological}%
  \BibitemOpen
  \bibfield  {author} {\bibinfo {author} {\bibfnamefont {J.~R.}\ \bibnamefont
  {Espinosa}} \emph {et~al.},\ }\href {\doibase 10.1103/PhysRevLett.115.251803}
  {\bibfield  {journal} {\bibinfo  {journal} {\emph{Phys. Rev. Lett.}}\
  }\textbf {\bibinfo {volume} {115}},\ \bibinfo {pages} {251803} (\bibinfo
  {year} {2015})}\BibitemShut {NoStop}%
\bibitem [{\citenamefont {Lazarides}\ and\ \citenamefont
  {Shafi}(2020)}]{LAZARIDES2020135603}%
  \BibitemOpen
  \bibfield  {author} {\bibinfo {author} {\bibfnamefont {G.}~\bibnamefont
  {Lazarides}}\ and\ \bibinfo {author} {\bibfnamefont {Q.}~\bibnamefont
  {Shafi}},\ }\href {\doibase https://doi.org/10.1016/j.physletb.2020.135603}
  {\bibfield  {journal} {\bibinfo  {journal} {\emph{Phys. Lett. B}}\ }\textbf
  {\bibinfo {volume} {807}},\ \bibinfo {pages} {135603} (\bibinfo {year}
  {2020})}\BibitemShut {NoStop}%
\bibitem [{\citenamefont {Bae}\ \emph {et~al.}(2015)\citenamefont {Bae},
  \citenamefont {Baer}, \citenamefont {Chun},\ and\ \citenamefont
  {Shin}}]{Bae2015Mixed}%
  \BibitemOpen
  \bibfield  {author} {\bibinfo {author} {\bibfnamefont {K.~J.}\ \bibnamefont
  {Bae}}, \bibinfo {author} {\bibfnamefont {H.}~\bibnamefont {Baer}}, \bibinfo
  {author} {\bibfnamefont {E.~J.}\ \bibnamefont {Chun}}, \ and\ \bibinfo
  {author} {\bibfnamefont {C.~S.}\ \bibnamefont {Shin}},\ }\href {\doibase
  10.1103/PhysRevD.91.075011} {\bibfield  {journal} {\bibinfo  {journal}
  {\emph{Phys. Rev. D}}\ }\textbf {\bibinfo {volume} {91}},\ \bibinfo {pages}
  {075011} (\bibinfo {year} {2015})}\BibitemShut {NoStop}%
\bibitem [{\citenamefont {Giannotti}\ \emph {et~al.}(2011)\citenamefont
  {Giannotti}, \citenamefont {Duffy},\ and\ \citenamefont
  {Nita}}]{Giannotti2011New}%
  \BibitemOpen
  \bibfield  {author} {\bibinfo {author} {\bibfnamefont {M.}~\bibnamefont
  {Giannotti}}, \bibinfo {author} {\bibfnamefont {L.}~\bibnamefont {Duffy}}, \
  and\ \bibinfo {author} {\bibfnamefont {R.}~\bibnamefont {Nita}},\ }\href
  {\doibase 10.1088/1475-7516/2011/01/015} {\bibfield  {journal} {\bibinfo
  {journal} {\emph{J. Cosmol. Astropart. Phys.}}\ }\textbf {\bibinfo {volume}
  {2011}},\ \bibinfo {pages} {015} (\bibinfo {year} {2011})}\BibitemShut
  {NoStop}%
\bibitem [{\citenamefont {Arias}\ \emph {et~al.}(2012)\citenamefont {Arias}
  \emph {et~al.}}]{Arias2012WISPy}%
  \BibitemOpen
  \bibfield  {author} {\bibinfo {author} {\bibfnamefont {P.}~\bibnamefont
  {Arias}} \emph {et~al.},\ }\href {\doibase 10.1088/1475-7516/2012/06/013}
  {\bibfield  {journal} {\bibinfo  {journal} {\emph{J. Cosmol. Astropart.
  Phys.}}\ }\textbf {\bibinfo {volume} {2012}},\ \bibinfo {pages} {013}
  (\bibinfo {year} {2012})}\BibitemShut {NoStop}%
\bibitem [{\citenamefont {Chavanis}(2021)}]{Chavanis2021Jeans}%
  \BibitemOpen
  \bibfield  {author} {\bibinfo {author} {\bibfnamefont {P.-H.}\ \bibnamefont
  {Chavanis}},\ }\href {https://doi.org/10.1103/PhysRevD.103.123551} {\bibfield
   {journal} {\bibinfo  {journal} {\emph{Phys. Rev. D}}\ }\textbf {\bibinfo
  {volume} {103}},\ \bibinfo {pages} {123551} (\bibinfo {year}
  {2021})}\BibitemShut {NoStop}%
\bibitem [{\citenamefont {Berezhiani}\ \emph {et~al.}(2021)\citenamefont
  {Berezhiani}, \citenamefont {Cintia},\ and\ \citenamefont
  {Warkentin}}]{BEREZHIANI2021Core}%
  \BibitemOpen
  \bibfield  {author} {\bibinfo {author} {\bibfnamefont {L.}~\bibnamefont
  {Berezhiani}}, \bibinfo {author} {\bibfnamefont {G.}~\bibnamefont {Cintia}},
  \ and\ \bibinfo {author} {\bibfnamefont {M.}~\bibnamefont {Warkentin}},\
  }\href {\doibase https://doi.org/10.1016/j.physletb.2021.136422} {\bibfield
  {journal} {\bibinfo  {journal} {\emph{Phys. Lett. B}}\ }\textbf {\bibinfo
  {volume} {819}},\ \bibinfo {pages} {136422} (\bibinfo {year}
  {2021})}\BibitemShut {NoStop}%
\bibitem [{\citenamefont {Levkov}\ \emph {et~al.}(2018)\citenamefont {Levkov},
  \citenamefont {Panin},\ and\ \citenamefont
  {Tkachev}}]{Levkov2018Gravitational}%
  \BibitemOpen
  \bibfield  {author} {\bibinfo {author} {\bibfnamefont {D.~G.}\ \bibnamefont
  {Levkov}}, \bibinfo {author} {\bibfnamefont {A.~G.}\ \bibnamefont {Panin}}, \
  and\ \bibinfo {author} {\bibfnamefont {I.~I.}\ \bibnamefont {Tkachev}},\
  }\href {\doibase 10.1103/physrevlett.121.151301} {\bibfield  {journal}
  {\bibinfo  {journal} {\emph{Phys. Rev. Lett.}}\ }\textbf {\bibinfo {volume}
  {121}},\ \bibinfo {pages} {151301} (\bibinfo {year} {2018})}\BibitemShut
  {NoStop}%
\bibitem [{\citenamefont {Destri}\ \emph {et~al.}(2013)\citenamefont {Destri},
  \citenamefont {{de Vega}},\ and\ \citenamefont
  {Sanchez}}]{Destri2013Quantum}%
  \BibitemOpen
  \bibfield  {author} {\bibinfo {author} {\bibfnamefont {C.}~\bibnamefont
  {Destri}}, \bibinfo {author} {\bibfnamefont {H.}~\bibnamefont {{de Vega}}}, \
  and\ \bibinfo {author} {\bibfnamefont {N.}~\bibnamefont {Sanchez}},\ }\href
  {\doibase https://doi.org/10.1016/j.astropartphys.2013.04.004} {\bibfield
  {journal} {\bibinfo  {journal} {\emph{Astropart. Phys.}}\ }\textbf {\bibinfo
  {volume} {46}},\ \bibinfo {pages} {14} (\bibinfo {year} {2013})}\BibitemShut
  {NoStop}%
\bibitem [{\citenamefont {de~Vega}\ \emph {et~al.}(2014)\citenamefont
  {de~Vega}, \citenamefont {Salucci},\ and\ \citenamefont
  {Sanchez}}]{deVega2014Observational}%
  \BibitemOpen
  \bibfield  {author} {\bibinfo {author} {\bibfnamefont {H.~J.}\ \bibnamefont
  {de~Vega}}, \bibinfo {author} {\bibfnamefont {P.}~\bibnamefont {Salucci}}, \
  and\ \bibinfo {author} {\bibfnamefont {N.~G.}\ \bibnamefont {Sanchez}},\
  }\href {\doibase 10.1093/mnras/stu972} {\bibfield  {journal} {\bibinfo
  {journal} {\emph{Mon. Not. R. Astron. Soc.}}\ }\textbf {\bibinfo {volume}
  {442}},\ \bibinfo {pages} {2717} (\bibinfo {year} {2014})}\BibitemShut
  {NoStop}%
\bibitem [{\citenamefont {de~Vega}\ and\ \citenamefont
  {Sanchez}(2017)}]{deVega2017Equation}%
  \BibitemOpen
  \bibfield  {author} {\bibinfo {author} {\bibfnamefont {H.~J.}\ \bibnamefont
  {de~Vega}}\ and\ \bibinfo {author} {\bibfnamefont {N.~G.}\ \bibnamefont
  {Sanchez}},\ }\href {https://doi.org/10.1140/epjc/s10052-017-4645-8}
  {\bibfield  {journal} {\bibinfo  {journal} {\emph{ Eur. Phys. J. C}}\
  }\textbf {\bibinfo {volume} {77}} (\bibinfo {year} {2017})}\BibitemShut
  {NoStop}%
\bibitem [{\citenamefont {Griffin}(1996)}]{Griffin1996Conserving}%
  \BibitemOpen
  \bibfield  {author} {\bibinfo {author} {\bibfnamefont {A.}~\bibnamefont
  {Griffin}},\ }\href {\doibase 10.1103/physrevb.53.9341} {\bibfield  {journal}
  {\bibinfo  {journal} {\emph{Phys. Rev. B}}\ }\textbf {\bibinfo {volume}
  {53}},\ \bibinfo {pages} {9341} (\bibinfo {year} {1996})}\BibitemShut
  {NoStop}%
\bibitem [{\citenamefont {Dalfovo}\ \emph {et~al.}(1999)\citenamefont
  {Dalfovo}, \citenamefont {Giorgini}, \citenamefont {Pitaevskii},\ and\
  \citenamefont {Stringari}}]{Dalfovo1999Theory}%
  \BibitemOpen
  \bibfield  {author} {\bibinfo {author} {\bibfnamefont {F.}~\bibnamefont
  {Dalfovo}}, \bibinfo {author} {\bibfnamefont {S.}~\bibnamefont {Giorgini}},
  \bibinfo {author} {\bibfnamefont {L.~P.}\ \bibnamefont {Pitaevskii}}, \ and\
  \bibinfo {author} {\bibfnamefont {S.}~\bibnamefont {Stringari}},\ }\href
  {\doibase 10.1103/revmodphys.71.463} {\bibfield  {journal} {\bibinfo
  {journal} {\emph{Rev. Mod. Phys.}}\ }\textbf {\bibinfo {volume} {71}},\
  \bibinfo {pages} {463} (\bibinfo {year} {1999})}\BibitemShut {NoStop}%
\bibitem [{\citenamefont {Pethick}\ and\ \citenamefont
  {Smith}(2008)}]{pethick_smith_2008}%
  \BibitemOpen
  \bibfield  {author} {\bibinfo {author} {\bibfnamefont {C.~J.}\ \bibnamefont
  {Pethick}}\ and\ \bibinfo {author} {\bibfnamefont {H.}~\bibnamefont
  {Smith}},\ }\href {\doibase 10.1017/CBO9780511802850} {\emph {\bibinfo
  {title} {Bose--Einstein Condensation in Dilute Gases}}},\ \bibinfo {edition}
  {2nd}\ ed.\ (\bibinfo  {publisher} {Cambridge University Press},\ \bibinfo
  {year} {2008})\BibitemShut {NoStop}%
\bibitem [{\citenamefont {Griffin}\ \emph {et~al.}(2009)\citenamefont
  {Griffin}, \citenamefont {Nikuni},\ and\ \citenamefont
  {Zaremba}}]{Griffin2009Bose}%
  \BibitemOpen
  \bibfield  {author} {\bibinfo {author} {\bibfnamefont {A.}~\bibnamefont
  {Griffin}}, \bibinfo {author} {\bibfnamefont {T.}~\bibnamefont {Nikuni}}, \
  and\ \bibinfo {author} {\bibfnamefont {E.}~\bibnamefont {Zaremba}},\ }\href
  {\doibase 10.1017/cbo9780511575150} {\emph {\bibinfo {title} {Bose-condensed
  gases at finite temperatures}}}\ (\bibinfo  {publisher} {Cambridge University
  Press},\ \bibinfo {year} {2009})\BibitemShut {NoStop}%
\bibitem [{\citenamefont {Yukalov}(2011)}]{Yukalov2011Basics}%
  \BibitemOpen
  \bibfield  {author} {\bibinfo {author} {\bibfnamefont {V.}~\bibnamefont
  {Yukalov}},\ }\href {\doibase 10.1134/s1063779611030063} {\bibfield
  {journal} {\bibinfo  {journal} {\emph{ Phys. Part. Nuclei}}\ }\textbf
  {\bibinfo {volume} {42}},\ \bibinfo {pages} {460} (\bibinfo {year}
  {2011})}\BibitemShut {NoStop}%
\bibitem [{\citenamefont {Chavanis}(2011)}]{Chavanis2011Mass}%
  \BibitemOpen
  \bibfield  {author} {\bibinfo {author} {\bibfnamefont {P.~H.}\ \bibnamefont
  {Chavanis}},\ }\href {\doibase 10.1103/physrevd.84.043531} {\bibfield
  {journal} {\bibinfo  {journal} {\emph{Phys. Rev. D}}\ }\textbf {\bibinfo
  {volume} {84}},\ \bibinfo {pages} {043531} (\bibinfo {year}
  {2011})}\BibitemShut {NoStop}%
\bibitem [{\citenamefont {Rindler-Daller}\ and\ \citenamefont
  {Shapiro}(2012)}]{Rindler-Daller2012Angular}%
  \BibitemOpen
  \bibfield  {author} {\bibinfo {author} {\bibfnamefont {T.}~\bibnamefont
  {Rindler-Daller}}\ and\ \bibinfo {author} {\bibfnamefont {P.~R.}\
  \bibnamefont {Shapiro}},\ }\href {\doibase 10.1111/j.1365-2966.2012.20588.x}
  {\bibfield  {journal} {\bibinfo  {journal} {\emph{Mon. Not. R. Astron.
  Soc.}}\ }\textbf {\bibinfo {volume} {422}},\ \bibinfo {pages} {135} (\bibinfo
  {year} {2012})}\BibitemShut {NoStop}%
\bibitem [{\citenamefont {Zhang}\ \emph {et~al.}(2018)\citenamefont {Zhang}
  \emph {et~al.}}]{Zhang2018Slowly}%
  \BibitemOpen
  \bibfield  {author} {\bibinfo {author} {\bibfnamefont {X.}~\bibnamefont
  {Zhang}} \emph {et~al.},\ }\href {\doibase 10.1140/epjc/s10052-018-5835-8}
  {\bibfield  {journal} {\bibinfo  {journal} {\emph{ Eur. Phys. J. C}}\
  }\textbf {\bibinfo {volume} {78}},\ \bibinfo {pages} {1} (\bibinfo {year}
  {2018})}\BibitemShut {NoStop}%
\bibitem [{\citenamefont {Guzm{\'a}n}\ \emph {et~al.}(2014)\citenamefont
  {Guzm{\'a}n}, \citenamefont {Lora-Clavijo}, \citenamefont
  {Gonz{\'a}lez-Avil{\'e}s},\ and\ \citenamefont
  {Rivera-Paleo}}]{Guzman2014Rotation}%
  \BibitemOpen
  \bibfield  {author} {\bibinfo {author} {\bibfnamefont {F.}~\bibnamefont
  {Guzm{\'a}n}}, \bibinfo {author} {\bibfnamefont {F.}~\bibnamefont
  {Lora-Clavijo}}, \bibinfo {author} {\bibfnamefont {J.}~\bibnamefont
  {Gonz{\'a}lez-Avil{\'e}s}}, \ and\ \bibinfo {author} {\bibfnamefont
  {F.}~\bibnamefont {Rivera-Paleo}},\ }\href {\doibase
  10.1103/physrevd.89.063507} {\bibfield  {journal} {\bibinfo  {journal}
  {\emph{Phys. Rev. D}}\ }\textbf {\bibinfo {volume} {89}},\ \bibinfo {pages}
  {063507} (\bibinfo {year} {2014})}\BibitemShut {NoStop}%
\bibitem [{\citenamefont {Hogan}\ and\ \citenamefont
  {Rees}(1988)}]{Hogan1988Axion}%
  \BibitemOpen
  \bibfield  {author} {\bibinfo {author} {\bibfnamefont {C.}~\bibnamefont
  {Hogan}}\ and\ \bibinfo {author} {\bibfnamefont {M.}~\bibnamefont {Rees}},\
  }\href {\doibase 10.1016/0370-2693(88)91655-3} {\bibfield  {journal}
  {\bibinfo  {journal} {\emph{Phys. Lett. B}}\ }\textbf {\bibinfo {volume}
  {205}},\ \bibinfo {pages} {228} (\bibinfo {year} {1988})}\BibitemShut
  {NoStop}%
\bibitem [{\citenamefont {Eggemeier}\ and\ \citenamefont
  {Niemeyer}(2019)}]{Eggemeier2019Formation}%
  \BibitemOpen
  \bibfield  {author} {\bibinfo {author} {\bibfnamefont {B.}~\bibnamefont
  {Eggemeier}}\ and\ \bibinfo {author} {\bibfnamefont {J.~C.}\ \bibnamefont
  {Niemeyer}},\ }\href {\doibase 10.1103/PhysRevD.100.063528} {\bibfield
  {journal} {\bibinfo  {journal} {\emph{Phys. Rev. D}}\ }\textbf {\bibinfo
  {volume} {100}},\ \bibinfo {pages} {063528} (\bibinfo {year}
  {2019})}\BibitemShut {NoStop}%
\bibitem [{\citenamefont {Sikivie}\ and\ \citenamefont
  {Yang}(2009)}]{Sikivie2009Bosea}%
  \BibitemOpen
  \bibfield  {author} {\bibinfo {author} {\bibfnamefont {P.}~\bibnamefont
  {Sikivie}}\ and\ \bibinfo {author} {\bibfnamefont {Q.}~\bibnamefont {Yang}},\
  }\href {https://link.aps.org/doi/10.1103/physrevlett.103.111301} {\bibfield
  {journal} {\bibinfo  {journal} {\emph{Phys. Rev. Lett.}}\ }\textbf {\bibinfo
  {volume} {103}} (\bibinfo {year} {2009})}\BibitemShut {NoStop}%
\bibitem [{\citenamefont {Mielczarek}\ \emph {et~al.}(2010)\citenamefont
  {Mielczarek}, \citenamefont {Stachowiak},\ and\ \citenamefont
  {Szyd{\l}owski}}]{Mielczarek2010Vortex}%
  \BibitemOpen
  \bibfield  {author} {\bibinfo {author} {\bibfnamefont {J.}~\bibnamefont
  {Mielczarek}}, \bibinfo {author} {\bibfnamefont {T.}~\bibnamefont
  {Stachowiak}}, \ and\ \bibinfo {author} {\bibfnamefont {M.}~\bibnamefont
  {Szyd{\l}owski}},\ }\href {\doibase 10.1142/s0218271810018037} {\bibfield
  {journal} {\bibinfo  {journal} {\emph{Int. J. Mod. Phys. D}}\ }\textbf
  {\bibinfo {volume} {19}},\ \bibinfo {pages} {1843} (\bibinfo {year}
  {2010})}\BibitemShut {NoStop}%
\bibitem [{\citenamefont {Tulin}\ and\ \citenamefont
  {Yu}(2018)}]{Tulin2018Dark}%
  \BibitemOpen
  \bibfield  {author} {\bibinfo {author} {\bibfnamefont {S.}~\bibnamefont
  {Tulin}}\ and\ \bibinfo {author} {\bibfnamefont {H.}~\bibnamefont {Yu}},\
  }\href {\doibase 10.1016/j.physrep.2017.11.004} {\bibfield  {journal}
  {\bibinfo  {journal} {\emph{Phys. Rep.}}\ }\textbf {\bibinfo {volume}
  {730}},\ \bibinfo {pages} {1} (\bibinfo {year} {2018})}\BibitemShut {NoStop}%
\bibitem [{\citenamefont {Mo}\ and\ \citenamefont {Mao}(2000)}]{Mo2000Tully}%
  \BibitemOpen
  \bibfield  {author} {\bibinfo {author} {\bibfnamefont {H.~J.}\ \bibnamefont
  {Mo}}\ and\ \bibinfo {author} {\bibfnamefont {S.}~\bibnamefont {Mao}},\
  }\href {\doibase 10.1046/j.1365-8711.2000.03714.x} {\bibfield  {journal}
  {\bibinfo  {journal} {\emph{Mon. Not. R. Astron. Soc.}}\ }\textbf {\bibinfo
  {volume} {318}},\ \bibinfo {pages} {163} (\bibinfo {year}
  {2000})}\BibitemShut {NoStop}%
\bibitem [{\citenamefont {Mitra}(2005)}]{Mitra2005Has}%
  \BibitemOpen
  \bibfield  {author} {\bibinfo {author} {\bibfnamefont {S.}~\bibnamefont
  {Mitra}},\ }\href {https://doi.org/10.1103/PhysRevD.71.121302} {\bibfield
  {journal} {\bibinfo  {journal} {\emph{Phys. Rev. D}}\ }\textbf {\bibinfo
  {volume} {71}} (\bibinfo {year} {2005})}\BibitemShut {NoStop}%
\bibitem [{\citenamefont {Bernal}\ \emph {et~al.}(2019)\citenamefont {Bernal},
  \citenamefont {Cosme},\ and\ \citenamefont
  {Tenkanen}}]{bernal2019phenomenology}%
  \BibitemOpen
  \bibfield  {author} {\bibinfo {author} {\bibfnamefont {N.}~\bibnamefont
  {Bernal}}, \bibinfo {author} {\bibfnamefont {C.}~\bibnamefont {Cosme}}, \
  and\ \bibinfo {author} {\bibfnamefont {T.}~\bibnamefont {Tenkanen}},\ }\href
  {https://doi.org/10.1140/epjc/s10052-019-6608-8} {\bibfield  {journal}
  {\bibinfo  {journal} {\emph{ Eur. Phys. J. C}}\ }\textbf {\bibinfo {volume}
  {79}},\ \bibinfo {pages} {99} (\bibinfo {year} {2019})}\BibitemShut {NoStop}%
\bibitem [{\citenamefont {Robertson}\ \emph {et~al.}(2019)\citenamefont
  {Robertson} \emph {et~al.}}]{Robertson2019mnras}%
  \BibitemOpen
  \bibfield  {author} {\bibinfo {author} {\bibfnamefont {A.}~\bibnamefont
  {Robertson}} \emph {et~al.},\ }\href {\doibase 10.1093/mnras/stz1815}
  {\bibfield  {journal} {\bibinfo  {journal} {\emph{Mon. Not. R. Astron.
  Soc.}}\ }\textbf {\bibinfo {volume} {488}},\ \bibinfo {pages} {3646}
  (\bibinfo {year} {2019})}\BibitemShut {NoStop}%
\bibitem [{\citenamefont {Randall}\ \emph {et~al.}(2008)\citenamefont {Randall}
  \emph {et~al.}}]{Randall2007Constraints}%
  \BibitemOpen
  \bibfield  {author} {\bibinfo {author} {\bibfnamefont {S.~W.}\ \bibnamefont
  {Randall}} \emph {et~al.},\ }\href {\doibase 10.1086/587859} {\bibfield
  {journal} {\bibinfo  {journal} {\emph{Astrophys. J.}}\ }\textbf {\bibinfo
  {volume} {679}},\ \bibinfo {pages} {1173} (\bibinfo {year}
  {2008})}\BibitemShut {NoStop}%
\bibitem [{\citenamefont {Brada{\v{c}}}\ \emph {et~al.}(2008)\citenamefont
  {Brada{\v{c}}} \emph {et~al.}}]{Bradac2008Revealing}%
  \BibitemOpen
  \bibfield  {author} {\bibinfo {author} {\bibfnamefont {M.}~\bibnamefont
  {Brada{\v{c}}}} \emph {et~al.},\ }\href {\doibase 10.1086/591246} {\bibfield
  {journal} {\bibinfo  {journal} {\emph{Astrophys. J.}}\ }\textbf {\bibinfo
  {volume} {687}},\ \bibinfo {pages} {959} (\bibinfo {year}
  {2008})}\BibitemShut {NoStop}%
\bibitem [{\citenamefont {Dwornik}\ \emph {et~al.}(2017)\citenamefont
  {Dwornik}, \citenamefont {Keresztes}, \citenamefont {Kun},\ and\
  \citenamefont {Gergely}}]{Dwornik2017Bose}%
  \BibitemOpen
  \bibfield  {author} {\bibinfo {author} {\bibfnamefont {M.}~\bibnamefont
  {Dwornik}}, \bibinfo {author} {\bibfnamefont {Z.}~\bibnamefont {Keresztes}},
  \bibinfo {author} {\bibfnamefont {E.}~\bibnamefont {Kun}}, \ and\ \bibinfo
  {author} {\bibfnamefont {L.}~\bibnamefont {Gergely}},\ }\href
  {https://doi.org/10.1155/2017/4025386} {\bibfield  {journal} {\bibinfo
  {journal} {\emph{Adv. High Energy Phys.}}\ }\textbf {\bibinfo {volume}
  {2017}} (\bibinfo {year} {2017})}\BibitemShut {NoStop}%
\bibitem [{\citenamefont {Harko}\ and\ \citenamefont
  {Madarassy}(2012)}]{Harko2012Finite}%
  \BibitemOpen
  \bibfield  {author} {\bibinfo {author} {\bibfnamefont {T.}~\bibnamefont
  {Harko}}\ and\ \bibinfo {author} {\bibfnamefont {E.~J.}\ \bibnamefont
  {Madarassy}},\ }\href {\doibase 10.1088/1475-7516/2012/01/020} {\bibfield
  {journal} {\bibinfo  {journal} {\emph{J. Cosmol. Astropart. Phys.}}\ }\textbf
  {\bibinfo {volume} {2012}},\ \bibinfo {pages} {020} (\bibinfo {year}
  {2012})}\BibitemShut {NoStop}%
\bibitem [{\citenamefont {Abdullin}\ and\ \citenamefont
  {Popov}(2019)}]{Abdullin2019Bose}%
  \BibitemOpen
  \bibfield  {author} {\bibinfo {author} {\bibfnamefont {I.~G.}\ \bibnamefont
  {Abdullin}}\ and\ \bibinfo {author} {\bibfnamefont {V.~A.}\ \bibnamefont
  {Popov}},\ }\href {\doibase 10.17238/issn2226-8812.2019.1.26-44} {\bibfield
  {journal} {\bibinfo  {journal} {\emph{Space, Time and Fundamental
  Interactions}}\ }\textbf {\bibinfo {volume} {1}},\ \bibinfo {pages} {26}
  (\bibinfo {year} {2019})}\BibitemShut {NoStop}%
\bibitem [{\citenamefont {Seidel}\ and\ \citenamefont
  {Suen}(1994)}]{Seidel1994Formation}%
  \BibitemOpen
  \bibfield  {author} {\bibinfo {author} {\bibfnamefont {E.}~\bibnamefont
  {Seidel}}\ and\ \bibinfo {author} {\bibfnamefont {W.-M.}\ \bibnamefont
  {Suen}},\ }\href {\doibase 10.1103/PhysRevLett.72.2516} {\bibfield  {journal}
  {\bibinfo  {journal} {\emph{Phys. Rev. Lett.}}\ }\textbf {\bibinfo {volume}
  {72}},\ \bibinfo {pages} {2516} (\bibinfo {year} {1994})}\BibitemShut
  {NoStop}%
\bibitem [{\citenamefont {Lynden-Bell}(1967)}]{Lynden-Bell1967Statistical}%
  \BibitemOpen
  \bibfield  {author} {\bibinfo {author} {\bibfnamefont {D.}~\bibnamefont
  {Lynden-Bell}},\ }\href {\doibase 10.1093/mnras/136.1.101} {\bibfield
  {journal} {\bibinfo  {journal} {\emph{Mon. Not. R. Astron. Soc.}}\ }\textbf
  {\bibinfo {volume} {136}},\ \bibinfo {pages} {101} (\bibinfo {year}
  {1967})}\BibitemShut {NoStop}%
\bibitem [{\citenamefont {Guzman}\ and\ \citenamefont
  {Urena-Lopez}(2006)}]{Guzman2006Gravitational}%
  \BibitemOpen
  \bibfield  {author} {\bibinfo {author} {\bibfnamefont {F.~S.}\ \bibnamefont
  {Guzman}}\ and\ \bibinfo {author} {\bibfnamefont {L.~A.}\ \bibnamefont
  {Urena-Lopez}},\ }\href {\doibase 10.1086/504508} {\bibfield  {journal}
  {\bibinfo  {journal} {\emph{Astrophys. J.}}\ }\textbf {\bibinfo {volume}
  {645}},\ \bibinfo {pages} {814} (\bibinfo {year} {2006})}\BibitemShut
  {NoStop}%
\bibitem [{\citenamefont {Chavanis}(2019)}]{Chavanis2019Predictive}%
  \BibitemOpen
  \bibfield  {author} {\bibinfo {author} {\bibfnamefont {P.-H.}\ \bibnamefont
  {Chavanis}},\ }\href {\doibase 10.1103/PhysRevD.100.083022} {\bibfield
  {journal} {\bibinfo  {journal} {\emph{Phys. Rev. D}}\ }\textbf {\bibinfo
  {volume} {100}},\ \bibinfo {pages} {083022} (\bibinfo {year}
  {2019})}\BibitemShut {NoStop}%
\bibitem [{\citenamefont {Kolb}\ and\ \citenamefont
  {Tkachev}(1994)}]{Kolb1994Large}%
  \BibitemOpen
  \bibfield  {author} {\bibinfo {author} {\bibfnamefont {E.~W.}\ \bibnamefont
  {Kolb}}\ and\ \bibinfo {author} {\bibfnamefont {I.~I.}\ \bibnamefont
  {Tkachev}},\ }\href {\doibase 10.1103/physrevd.50.769} {\bibfield  {journal}
  {\bibinfo  {journal} {\emph{Phys. Rev. D}}\ }\textbf {\bibinfo {volume}
  {50}},\ \bibinfo {pages} {769} (\bibinfo {year} {1994})}\BibitemShut
  {NoStop}%
\bibitem [{\citenamefont {Berezinsky}\ \emph {et~al.}(2003)\citenamefont
  {Berezinsky}, \citenamefont {Dokuchaev},\ and\ \citenamefont
  {Eroshenko}}]{Berezinsky2003Small}%
  \BibitemOpen
  \bibfield  {author} {\bibinfo {author} {\bibfnamefont {V.}~\bibnamefont
  {Berezinsky}}, \bibinfo {author} {\bibfnamefont {V.}~\bibnamefont
  {Dokuchaev}}, \ and\ \bibinfo {author} {\bibfnamefont {Y.}~\bibnamefont
  {Eroshenko}},\ }\href {\doibase 10.1103/PhysRevD.68.103003} {\bibfield
  {journal} {\bibinfo  {journal} {\emph{Phys. Rev. D}}\ }\textbf {\bibinfo
  {volume} {68}},\ \bibinfo {pages} {103003} (\bibinfo {year}
  {2003})}\BibitemShut {NoStop}%
\bibitem [{\citenamefont {Breit}\ \emph {et~al.}(1984)\citenamefont {Breit},
  \citenamefont {Gupta},\ and\ \citenamefont {Zaks}}]{Breit1984Cold}%
  \BibitemOpen
  \bibfield  {author} {\bibinfo {author} {\bibfnamefont {J.}~\bibnamefont
  {Breit}}, \bibinfo {author} {\bibfnamefont {S.}~\bibnamefont {Gupta}}, \ and\
  \bibinfo {author} {\bibfnamefont {A.}~\bibnamefont {Zaks}},\ }\href {\doibase
  https://doi.org/10.1016/0370-2693(84)90764-0} {\bibfield  {journal} {\bibinfo
   {journal} {\emph{Phys. Lett. B}}\ }\textbf {\bibinfo {volume} {140}},\
  \bibinfo {pages} {329} (\bibinfo {year} {1984})}\BibitemShut {NoStop}%
\bibitem [{\citenamefont {Bergstr\"om}\ \emph {et~al.}(1999)\citenamefont
  {Bergstr\"om}, \citenamefont {Edsj\"o}, \citenamefont {Gondolo},\ and\
  \citenamefont {Ullio}}]{Bergstrom1999Clumpy}%
  \BibitemOpen
  \bibfield  {author} {\bibinfo {author} {\bibfnamefont {L.}~\bibnamefont
  {Bergstr\"om}}, \bibinfo {author} {\bibfnamefont {J.}~\bibnamefont
  {Edsj\"o}}, \bibinfo {author} {\bibfnamefont {P.}~\bibnamefont {Gondolo}}, \
  and\ \bibinfo {author} {\bibfnamefont {P.}~\bibnamefont {Ullio}},\ }\href
  {\doibase 10.1103/PhysRevD.59.043506} {\bibfield  {journal} {\bibinfo
  {journal} {\emph{Phys. Rev. D}}\ }\textbf {\bibinfo {volume} {59}},\ \bibinfo
  {pages} {043506} (\bibinfo {year} {1999})}\BibitemShut {NoStop}%
\bibitem [{\citenamefont {Binney}\ and\ \citenamefont
  {Tremaine}(2011)}]{binney2011galactic}%
  \BibitemOpen
  \bibfield  {author} {\bibinfo {author} {\bibfnamefont {J.}~\bibnamefont
  {Binney}}\ and\ \bibinfo {author} {\bibfnamefont {S.}~\bibnamefont
  {Tremaine}},\ }\href {https://ui.adsabs.harvard.edu/abs/1987gady.book.....B}
  {\emph {\bibinfo {title} {Galactic dynamics}}}\ (\bibinfo  {publisher}
  {Princeton university press},\ \bibinfo {year} {2011})\BibitemShut {NoStop}%
\bibitem [{\citenamefont {Boyanovsky}\ \emph {et~al.}(2008)\citenamefont
  {Boyanovsky}, \citenamefont {de~Vega},\ and\ \citenamefont
  {Sanchez}}]{Boyanovsky2008Constraints}%
  \BibitemOpen
  \bibfield  {author} {\bibinfo {author} {\bibfnamefont {D.}~\bibnamefont
  {Boyanovsky}}, \bibinfo {author} {\bibfnamefont {H.~J.}\ \bibnamefont
  {de~Vega}}, \ and\ \bibinfo {author} {\bibfnamefont {N.~G.}\ \bibnamefont
  {Sanchez}},\ }\href {\doibase 10.1103/PhysRevD.77.043518} {\bibfield
  {journal} {\bibinfo  {journal} {\emph{Phys. Rev. D}}\ }\textbf {\bibinfo
  {volume} {77}},\ \bibinfo {pages} {043518} (\bibinfo {year}
  {2008})}\BibitemShut {NoStop}%
\bibitem [{\citenamefont {Tremaine}\ and\ \citenamefont
  {Gunn}(1979)}]{Tremaine1979Dynamical}%
  \BibitemOpen
  \bibfield  {author} {\bibinfo {author} {\bibfnamefont {S.}~\bibnamefont
  {Tremaine}}\ and\ \bibinfo {author} {\bibfnamefont {J.~E.}\ \bibnamefont
  {Gunn}},\ }\href {\doibase 10.1103/PhysRevLett.42.407} {\bibfield  {journal}
  {\bibinfo  {journal} {\emph{Phys. Rev. Lett.}}\ }\textbf {\bibinfo {volume}
  {42}},\ \bibinfo {pages} {407} (\bibinfo {year} {1979})}\BibitemShut
  {NoStop}%
\bibitem [{\citenamefont {Madsen}(1990)}]{Madsen1990Phase}%
  \BibitemOpen
  \bibfield  {author} {\bibinfo {author} {\bibfnamefont {J.}~\bibnamefont
  {Madsen}},\ }\href {\doibase 10.1103/PhysRevLett.64.2744} {\bibfield
  {journal} {\bibinfo  {journal} {\emph{Phys. Rev. Lett.}}\ }\textbf {\bibinfo
  {volume} {64}},\ \bibinfo {pages} {2744} (\bibinfo {year}
  {1990})}\BibitemShut {NoStop}%
\bibitem [{\citenamefont {Madsen}(1991)}]{Madsen1991Generalized}%
  \BibitemOpen
  \bibfield  {author} {\bibinfo {author} {\bibfnamefont {J.}~\bibnamefont
  {Madsen}},\ }\href {\doibase 10.1103/PhysRevD.44.999} {\bibfield  {journal}
  {\bibinfo  {journal} {\emph{Phys. Rev. D}}\ }\textbf {\bibinfo {volume}
  {44}},\ \bibinfo {pages} {999} (\bibinfo {year} {1991})}\BibitemShut
  {NoStop}%
\bibitem [{\citenamefont {Simon}\ \emph {et~al.}(2011)\citenamefont {Simon}
  \emph {et~al.}}]{Simon2011COMPLETE}%
  \BibitemOpen
  \bibfield  {author} {\bibinfo {author} {\bibfnamefont {J.~D.}\ \bibnamefont
  {Simon}} \emph {et~al.},\ }\href {\doibase 10.1088/0004-637x/733/1/46}
  {\bibfield  {journal} {\bibinfo  {journal} {\emph{Astrophys. J.}}\ }\textbf
  {\bibinfo {volume} {733}},\ \bibinfo {pages} {46} (\bibinfo {year}
  {2011})}\BibitemShut {NoStop}%
\bibitem [{\citenamefont {Willman}\ and\ \citenamefont
  {Strader}(2012)}]{Willman2012GALAXY}%
  \BibitemOpen
  \bibfield  {author} {\bibinfo {author} {\bibfnamefont {B.}~\bibnamefont
  {Willman}}\ and\ \bibinfo {author} {\bibfnamefont {J.}~\bibnamefont
  {Strader}},\ }\href {\doibase 10.1088/0004-6256/144/3/76} {\bibfield
  {journal} {\bibinfo  {journal} {\emph{Astron. J.}}\ }\textbf {\bibinfo
  {volume} {144}},\ \bibinfo {pages} {76} (\bibinfo {year} {2012})}\BibitemShut
  {NoStop}%
\bibitem [{\citenamefont {de~Blok}\ and\ \citenamefont
  {McGaugh}(1997)}]{Blok1997dark}%
  \BibitemOpen
  \bibfield  {author} {\bibinfo {author} {\bibfnamefont {W.~J.~G.}\
  \bibnamefont {de~Blok}}\ and\ \bibinfo {author} {\bibfnamefont {S.~S.}\
  \bibnamefont {McGaugh}},\ }\href {\doibase 10.1093/mnras/290.3.533}
  {\bibfield  {journal} {\bibinfo  {journal} {\emph{Mon. Not. R. Astron.
  Soc.}}\ }\textbf {\bibinfo {volume} {290}},\ \bibinfo {pages} {533} (\bibinfo
  {year} {1997})}\BibitemShut {NoStop}%
\bibitem [{\citenamefont {McGaugh}\ \emph {et~al.}(2001)\citenamefont
  {McGaugh}, \citenamefont {Rubin},\ and\ \citenamefont
  {de~Blok}}]{McGaugh2001High}%
  \BibitemOpen
  \bibfield  {author} {\bibinfo {author} {\bibfnamefont {S.~S.}\ \bibnamefont
  {McGaugh}}, \bibinfo {author} {\bibfnamefont {V.~C.}\ \bibnamefont {Rubin}},
  \ and\ \bibinfo {author} {\bibfnamefont {W.~J.~G.}\ \bibnamefont {de~Blok}},\
  }\href {\doibase 10.1086/323448} {\bibfield  {journal} {\bibinfo  {journal}
  {\emph{Astron. J.}}\ }\textbf {\bibinfo {volume} {122}},\ \bibinfo {pages}
  {2381} (\bibinfo {year} {2001})}\BibitemShut {NoStop}%
\bibitem [{\citenamefont {Burkert}(1995)}]{Burkert1995Structure}%
  \BibitemOpen
  \bibfield  {author} {\bibinfo {author} {\bibfnamefont {A.}~\bibnamefont
  {Burkert}},\ }\href {\doibase 10.1007/978-94-009-0229-9_25} {\bibfield
  {journal} {\bibinfo  {journal} {\emph{Astrophys. J. Lett.}}\ }\textbf
  {\bibinfo {volume} {447}},\ \bibinfo {pages} {L25} (\bibinfo {year}
  {1995})}\BibitemShut {NoStop}%
\bibitem [{\citenamefont {Navarro}\ \emph {et~al.}(1996)\citenamefont
  {Navarro}, \citenamefont {Frenk},\ and\ \citenamefont
  {White}}]{Navarro1995Structure}%
  \BibitemOpen
  \bibfield  {author} {\bibinfo {author} {\bibfnamefont {J.~F.}\ \bibnamefont
  {Navarro}}, \bibinfo {author} {\bibfnamefont {C.~S.}\ \bibnamefont {Frenk}},
  \ and\ \bibinfo {author} {\bibfnamefont {S.~D.~M.}\ \bibnamefont {White}},\
  }\href {\doibase 10.1086/177173} {\bibfield  {journal} {\bibinfo  {journal}
  {\emph{Astrophys. J.}}\ }\textbf {\bibinfo {volume} {462}},\ \bibinfo {pages}
  {563} (\bibinfo {year} {1996})}\BibitemShut {NoStop}%
\end{thebibliography}%

\end{document}